\documentclass[12pt]{article}
\usepackage[T1]{fontenc} 

\usepackage{amsmath, amsfonts, amssymb} 
\usepackage{chemformula} 

\usepackage[colorlinks, linkcolor=black, anchorcolor=black, citecolor=black]{hyperref} 
\usepackage{graphicx} 
\usepackage{booktabs}
\usepackage{multirow} 
\usepackage{caption} 
\usepackage{subcaption} 
\usepackage{float} 
\usepackage{algorithm} 
\usepackage{booktabs}
\usepackage{makecell}
\usepackage{pifont}
\usepackage{authblk}
\usepackage{algpseudocode} 
\usepackage{enumitem} 
\usepackage{xcolor} 
\usepackage{adjustbox} 
\usepackage{natbib} 

\newtheorem{theorem}{Theorem}

\usepackage{siunitx}
\begin{document}
\title{\textbf{Deep Learning Enhanced Multivariate GARCH}}
\author{Haoyuan Wang}
\author{Chen Liu}
\author{Minh-Ngoc Tran}
\author{Chao Wang\thanks{Corresponding author: chao.wang@sydney.edu.au.}}

\affil{Discipline of Business Analytics, The University of Sydney Business School}
\date{}
\maketitle


\begin{abstract}
This paper introduces a novel multivariate volatility modeling framework, named Long Short-Term Memory enhanced BEKK (LSTM-BEKK), that integrates deep learning into multivariate GARCH processes. By combining the flexibility of recurrent neural networks with the econometric structure of BEKK models, our approach is designed to better capture nonlinear, dynamic, and high-dimensional dependence structures in financial return data. The proposed model addresses key limitations of traditional multivariate GARCH-based methods, particularly in capturing persistent volatility clustering and asymmetric co-movement across assets. Leveraging the data-driven nature of LSTMs, the framework adapts effectively to time-varying market conditions, offering improved robustness and forecasting performance. Empirical results across multiple equity markets confirm that the LSTM-BEKK model achieves superior performance in terms of out-of-sample portfolio risk forecast, while maintaining the interpretability from the BEKK models. These findings highlight the potential of hybrid econometric-deep learning models in advancing financial risk management and multivariate volatility forecasting.
\end{abstract}

\vspace{0.5em}

\newpage
    \section{Introduction}

Modeling financial market volatility has long been a central topic in econometrics due to its critical role in risk management, asset pricing, and portfolio optimization. \citet{Engle1982} pioneered this line of research with the introduction of the Autoregressive Conditional Heteroskedasticity model, which characterizes time-varying volatility as a function of past shocks. This foundational framework was later generalized by \citet{Bollerslev1986} through the GARCH model, which incorporated both lagged innovations and past variances, enabling improved modeling of persistent volatility behavior. In parallel to the GARCH family, stochastic volatility (SV) models emerged as an important alternative, modeling volatility as an unobserved latent process governed by its own stochastic dynamics. This latent-state formulation allows SV models to capture the stylized facts often observed in financial time series \citep{Taylor1994, Asai2006}.

While univariate volatility models are effective in capturing the dynamics of individual asset volatility, financial markets are inherently multivariate, with assets exhibiting strong comovements and spillovers. Accurate modeling of such joint dynamics is essential for systemic risk monitoring, portfolio allocation, and derivative pricing. In this context, the conditional covariance matrix plays a central role by describing the time-varying co-movement among asset returns. As highlighted in \citet{Bollerslev1988}, \citet{Engle2002} and \citet{Bauwens2006}, multivariate volatility modeling enables the quantification of interdependencies across assets and enhances the effectiveness of financial decision-making under uncertainty.

To extend volatility modeling to multivariate settings, multivariate GARCH (MGARCH) models have been proposed as natural generalizations of the univariate framework. Among these, the BEKK model introduced by \citet{EngleKroner1995} stands out due to its flexible parameterization and being a direct extension of univariate GARCH. The BEKK specification guarantees the positive definiteness of the conditional covariance matrix by construction and can capture dynamic spillovers between asset returns. This structural advantage makes it particularly appealing for applications requiring robust covariance estimation in financial risk modeling and forecasting.
Empirical studies \citep[see, e.g.,][]{SilvennoinenTerasvirta2009,Fang2015} have demonstrated the adaptability of MGARCH frameworks in capturing time-varying correlations and volatilities under extreme events, reinforcing their relevance in financial risk management. 

Traditional MGARCH formulations such as BEKK encounter severe computational bottlenecks in high-dimensional settings due to the rapidly expanding parameter space \citep{LedoitWolf2012, LedoitWolf2015}. To address this scalability issue, \citet{Engle2002} proposed the Dynamic Conditional Correlation (DCC) model, which simplifies estimation by decoupling univariate volatility and correlation dynamics. This reduction in complexity allows DCC to accommodate a larger number of assets while still capturing time-varying dependencies. Extensions such as the Student-\textit{t} DCC \citep{Ku2008} and Asymmetric DCC \citep{Lai2011} further enhance the modeling robustness under heavy tails and asymmetric shocks, while the Dynamic Equicorrelation model \citep{EngleKelly2012} further improves computational tractability for the DCC framework. 

These methodological contributions collectively underpin the development of modern multivariate volatility modeling, which is central to understanding the dynamic behavior of financial markets and the evolving interdependencies among assets. While models such as BEKK, DCC and their extensions have significantly advanced the multivariate volatility modeling, their reliance on the simple summation of lagged covariance matrices and outer products of return vectors limits their adaptability to complicated patterns and structural shifts often observed during periods of financial stress. These limitations motivate the integration of more expressive modeling techniques, such as deep learning models, into multivariate volatility modeling, offering potential new directions for advancing volatility modeling in high-dimensional financial applications.


Recent advances in deep learning provide compelling techniques for modeling complicated and high dimensional sequential data. Recurrent Neural Networks (RNNs), and Long Short-Term Memory (LSTM) networks in particular, offer a powerful mechanism for learning temporal and comovement dependencies in sequential multivariate data. Their ability to capture long-range interactions and nonlinear patterns has led to various successful applications across a range of large-scale industry level tasks \citep{Goodfellow2016}. 
This paper proposes a novel hybrid model, the LSTM-BEKK, which combines the econometric rigor of the BEKK model with the adaptive capabilities of LSTMs. The LSTM-BEKK model leverages the structural strengths of BEKK while enhancing it with LSTM's ability to capture nonlinear and long-term dynamics. By allowing LSTM-generated components to directly influence the time-varying covariance matrix, this framework provides a highly flexible tool for analyzing the evolving relationships among financial assets, particularly in high-dimensional contexts.

The design of the LSTM-BEKK model is inspired by recent advancements in integrating deep learning with univariate volatility models, which have demonstrated superior predictive accuracy and the ability to capture nonlinearities in volatility. For example, the work by \citet{Nguyen2022} explored the effectiveness of deep learning in enhancing GARCH-based volatility modeling. These studies highlight the advantages of incorporating neural networks into traditional econometric frameworks, improving the adaptability and forecasting performance of volatility models. Extending this approach to a multivariate setting introduces unique challenges, such as ensuring positive definiteness of the covariance matrix and managing the curse of dimensionality \citep{LedoitWolf2012}.
The primary innovation of the LSTM-BEKK model lies in its ability to utilize economic information instruments and adapt to changing market conditions. Unlike traditional BEKK models constrained by their inflexible parametric structure, the LSTM-BEKK model dynamically adjusts itself to capture evolving market relationships. Furthermore, the LSTM's capacity to learn complex patterns enhances the model's responsiveness to turbulent market periods, improving the accuracy of volatility forecasts.

Substantial empirical results demonstrate that the LSTM-BEKK model outperforms traditional BEKK and DCC models in terms of predictive accuracy, as measured by standard evaluation metrics such as out-of-sample negative log-likelihood and annualized volatility of global minimum variance portfolio. These findings are robust across datasets covering portfolios constructed from the top companies by market capitalization in Japan, the U.S., and the U.K., reflecting the model's strong generalization capability across different market environments. Moreover, in low-dimensional settings, the interpretability of the LSTM-BEKK framework is enhanced by visualizing individual variance and covariance trajectories, which reveal the model's ability to capture both abrupt volatility spikes and directional shifts in inter-asset correlations—especially during periods of market stress. These insights affirm not only its predictive power but also its value in understanding the evolving structure of financial return dynamics.

The paper is organized as follows. In Section \ref{sec:modelling_framework}, we review the relevant literature on MGARCH models and machine learning techniques, then detail the structure and theoretical foundations of the proposed LSTM-BEKK model. The estimation procedure of the LSTM-BEKK is detailed in Section \ref{sec:estimation}. Section \ref{sec:empirical_study} evaluates the empirical performance of LSTM-BEKK on high-dimensional datasets across multiple settings, comparing it with established benchmarks. A study focusing on global minimum variance portfolio is conducted in Section \ref{sec:portfolio}. Section \ref{sec:conclusion} concludes the paper and discusses future work. Technical details and further empirical study of LSTM-BEKK are included in the Appendix.  

\section{Modeling Frameworks}\label{sec:modelling_framework}

\subsection{Foundation Models}

This section presents the foundation models from econometrics and machine learning that form the building blocks for our proposed LSTM-BEKK model. We also present several benchmark multivariate volatility models that will be used to compare against LSTM-BEKK.

\subsubsection{BEKK Models}

The BEKK model is a representative within the MGARCH framework, designed to guarantee the positive definiteness of the conditional covariance matrix while preserving flexibility in modeling dynamic dependencies across financial assets. 

Let $\mathbf{r}_t = (\mathbf{r}_{t,1}, \dots, \mathbf{r}_{t,n})'$ denote the vector of de-meaned returns for $n$ portfolio assets at time $t$. The returns are assumed to follow a multivariate normal distribution conditional on past information $\mathcal{F}_{t-1}$:
\begin{equation} \label{eq:multivariate_normal}
    \mathbf{r}_t  | \mathcal{F}_{t-1} \sim N(\mathbf{0}, \mathbf{H}_t),
\end{equation}
where $\mathbf{H}_t = \text{cov}(\mathbf{r}_t | \mathcal{F}_{t-1})$ represents the conditional covariance matrix of returns.
While it is possible to consider more flat-tailed distributions such as a multivariate Student's $t$, we use the multivariate normal distribution in this paper to facilitate exposition.
This covariance matrix $\mathbf{H}_t$ captures time-varying dependencies among portfolio assets, a critical element for financial applications, such as risk management and portfolio allocation \citep{Bollerslev1988, Bauwens2006, McAleerChanHotiLieberman2008}.

The general BEKK($p,q$) model specifies $\mathbf{H}_t$ as:
\begin{equation}
    \mathbf{H}_t = \mathbf{\Omega} + \sum_{i=1}^{p} \mathbf{A}_i \mathbf{r}_{t-i} \mathbf{r}_{t-i}' \mathbf{A}_i' + \sum_{j=1}^{q} \mathbf{B}_j \mathbf{H}_{t-j} \mathbf{B}_j',
\end{equation}
where $\mathbf{\Omega}$ is a symmetric positive definite matrix, and $\mathbf{A}_i$ and $\mathbf{B}_j$ are $n \times n$ coefficient matrices capturing the effects of past shocks and past covariances, and $p$ and $q$ represent the orders of the process \citep{FrancqZakoian2012, ScherrerRibarits2007}. To reduce complexity, the BEKK(1,1) model is commonly used, which assumes $p = q = 1$, leading to the formulation:
\begin{equation}
    \mathbf{H}_t = \mathbf{\Omega} +  \mathbf{A}_1 \mathbf{r}_{t-1} \mathbf{r}_{t-1}' \mathbf{A}_1' + \mathbf{B}_1 \mathbf{H}_{t-1} \mathbf{B}_1'.
\end{equation}

A further simplification of the BEKK model is the Scalar BEKK specification, which simplifies the parameterization by imposing the following constraints: $\mathbf{\Omega} = \mathbf{C}\mathbf{C}'$, $\mathbf{A}_1 = \sqrt{a} \mathbf{I}$, and $\mathbf{B}_1 = \sqrt{b} \mathbf{I}$, where $\mathbf{C}$ is a lower triangular matrix and $\mathbf{I}$ denotes the identity matrix. This results in the more compact form
$\mathbf{H}_t$ as:
\begin{equation}
    \mathbf{H}_t = \mathbf{C}\mathbf{C}' + a \mathbf{r}_{t-1}\mathbf{r}_{t-1}' + b \mathbf{H}_{t-1}.
\end{equation}
Here, $a, b \geq 0$ are scalar parameters representing the effects of past shocks and lagged covariances, respectively.
The diagonal elements of $\mathbf{C}$ are assumed to be strictly non-zero, guaranteeing the positive definiteness of $\mathbf{\Omega} = \mathbf{C}\mathbf{C}'$, hence $\mathbf{H}_t$ \citep{Francq2019, MatsuiPedersen2022, HafnerPreminger2009}.
Moreover, the stationarity condition $a + b < 1$ ensures the decay of volatility over time, preserving the long-run stability of the process \citep{ScherrerRibarits2007, HafnerLaurentViolante2017}. This parsimonious structure drastically reduces the number of parameters in full BEKK models.

From an econometric perspective, the parameters $a$ and $b$ have intuitive interpretations. The parameter $a$ measures the sensitivity of the conditional covariance matrix to the past shocks, capturing the immediate effect of return innovations on volatility. Meanwhile, $b$ reflects the persistence of volatility over time, characterizing how past volatilities influence future dynamics. Together, these parameters provide a framework for understanding how risk propagates through time in financial markets.

The Scalar BEKK model postulates the conditional covariance matrix as a simple linear summation between its lagged value and the outer product of the past shock vector. While this assumption facilitates estimation and ensures computational tractability, it limits the model’s ability to capture more complex, nonlinear relationships among assets \citep{Caporin2008, HafnerRombouts2007, ScherrerRibarits2007}. These limitations inspire the development of our LSTM-BEKK
model, presented in Section \ref{subsec:lstmbekkmodel}, which integrates machine learning techniques into Scalar BEKK to enhance its adaptability and ability to capture richer dynamics, while still maintaining its 
econometric interpretability. 

In summary, the Scalar BEKK model strikes a balance between simplicity and efficacy, making it an essential tool for modeling multivariate volatility in financial markets. Its parsimonious structure and intuitive economic interpretation make it particularly suitable for applications such as portfolio risk management, systemic risk analysis, and stress testing. However, its reliance on over-parsimonious  parameterization necessitates further extensions, to address the complexities of real-world financial data.

\subsubsection{Dynamic Conditional Correlation (DCC) Model}

The DCC model, introduced by Engle \citeyearpar{Engle2002}, represents a major advancement in multivariate volatility modeling by efficiently capturing time-varying correlations in high-dimensional datasets. Unlike the full BEKK model, which suffers from parameter proliferation in large systems, the DCC model decomposes the conditional covariance matrix into conditional variances and correlations, enabling a computationally efficient estimation framework. The DCC model decomposes $\mathbf{H}_t$ as:
\begin{equation}
    \mathbf{H}_t = \mathbf{D}_t \mathbf{R}_t \mathbf{D}_t,
\end{equation}
where $\mathbf{D}_t = \text{diag}(\sqrt{h_{t,1}}, \dots, \sqrt{h_{t,n}})$ is a diagonal matrix of conditional standard deviations, and $\mathbf{R}_t$ is the conditional correlation matrix.

The diagonal elements of $\mathbf{D}_t$ are modeled as univariate GARCH processes:
\begin{equation}
    h_{i,t} = \omega_i + \alpha_i r_{i,t-1}^2 + \beta_i h_{i,t-1}, \quad i = 1, \dots, n,
\end{equation}
where $\omega_i > 0$, $\alpha_i \geq 0$, and $\beta_i \geq 0$ and $\alpha_i+\beta_i<1$ ensure positivity and stationarity of the conditional variances. The correlation dynamics are governed by the intermediate matrix $\mathbf{Q}_t$, updated recursively as:
\begin{equation}
    \mathbf{Q}_t = (1 - a - b)\mathbf{S} + a \mathbf{z}_{t-1} \mathbf{z}_{t-1}' + b\mathbf{Q}_{t-1},
\end{equation}
where $\mathbf{z}_t = \mathbf{D}_t^{-1} \mathbf{r}_t$ is the vector of standardized residuals, and $\mathbf{S}$ is the unconditional covariance matrix of $\mathbf{z}_t$. In this formulation, the parameters \( a \) and \( b \) play central roles in determining the dynamics of the conditional correlation matrix. Specifically, \( a \) governs the sensitivity of correlations to recent shocks in the standardized residuals (i.e., the innovation effect), while \( b \) controls the persistence of past correlations. The sum \( a + b < 1 \) is imposed to ensure stationarity and to guarantee that the conditional correlation matrix remains well-defined over time. Together, these parameters dictate the responsiveness and memory of the correlation dynamics, with higher values indicating stronger persistence and slower adaptation to new information. The matrix $\mathbf{S}$ is typically estimated as the sample covariance of the standardized residuals during the initial estimation stage:
\begin{equation}
    \mathbf{S} = \frac{1}{T} \sum_{t=1}^T \mathbf{z}_t \mathbf{z}_t',
\end{equation}
where $T$ is the sample size. The conditional correlation matrix $\mathbf{R}_t$ is obtained by standardizing $\mathbf{Q}_t$:
\begin{equation}
    \mathbf{R}_t = \text{diag}(\mathbf{Q}_t)^{-1/2} \mathbf{Q}_t \text{diag}(\mathbf{Q}_t)^{-1/2}.
\end{equation}
This ensures that $\mathbf{R}_t$ is symmetric, positive definite, and has unit diagonal elements.


The modular structure of the DCC model facilitates efficient estimation. In the first step, univariate GARCH models are estimated to compute $\mathbf{D}_t$. In the second step, the correlation dynamics are estimated using the standardized residuals $\mathbf{z}_t$. This separation reduces computational complexity compared to fully parameterized MGARCH models \citep{FrancqZakoian2012, Bauwens2006}, making the DCC model scalable to high-dimensional datasets.
The DCC model has been widely adopted in applications such as portfolio optimization, risk management, and systemic risk evaluation \citep{Bauwens2006, EngleKelly2012}. 
Extensions such as the corrected DCC model of \citet{Aielli2013} and regime-switching DCC variants of \citet{BAUWENS2020} further improve the adaptability of DCC in different application contexts. 

Overall, the DCC model represents a foundational tool in multivariate volatility modeling due to its balance between parsimony and effectiveness. However, evolving market complexity increasingly demands more expressive models. Innovations such as deep learning-augmented structures offer promising pathways to improve upon traditional frameworks and better accommodate the intricacies of modern asset return dynamics.

\subsubsection{LSTM Model}

Deep learning methods, particularly RNNs, have proven to be powerful tools for modeling sequential multivariate data. Among these, the LSTM network, introduced by Hochreiter and Schmidhuber \citeyearpar{Hochreiter1997}, stands out to be one of the most commonly used and effective RNN models. Its gating mechanism enables the selective retention of long-term dependencies, making it particularly suitable for applications in time series analysis, including financial volatility modeling \citep{Goodfellow2016}. The architecture of LSTM networks comprises three primary gates: the input gate ($g_{it}$), forget gate ($g_{ft}$), and output gate ($g_{ot}$). These gates regulate the flow of information, dynamically updating the cell state ($c_t$) and hidden state ($h_t$) to capture long-term and short-term patterns. 
Let $x_t$ be the input vector at time $t$, and $y_t$ be the forecast variable of interest.   
The evolution of these components can be described by the following equations:
\begin{subequations} \label{eq:lstm_updates}
    \begin{align}
        g_{it} &= \sigma \left( W_i [h_{t-1}, x_t] + b_i \right), \label{eq:lstm_input} \\
        g_{ft} &= \sigma \left( W_f [h_{t-1}, x_t] + b_f \right), \label{eq:lstm_forget} \\
        g_{ot} &= \sigma \left( W_o [h_{t-1}, x_t] + b_o \right), \label{eq:lstm_output} \\
        \tilde{c}_t &= \tanh \left( W_c [h_{t-1}, x_t] + b_c \right), \label{eq:lstm_candidate} \\
        c_t &= g_{ft} \odot c_{t-1} + g_{it} \odot \tilde{c}_t, \label{eq:lstm_cell} \\
        h_t &= g_{ot} \odot \tanh(c_t), \label{eq:lstm_hidden}
    \end{align}
\end{subequations}
where $\sigma(\cdot)$ is the sigmoid activation function, $\tanh(\cdot)$ is the hyperbolic tangent function, and $\odot$ denotes element-wise multiplication. 
The hidden state $h_t$ is then linked to the forecast variable $y_t$ by a measurement equation specified depending on the application context. 
These equations allow the LSTM network to adaptively learn from sequential data by managing the flow of information across time steps.
The reader is referred to \citet{Goodfellow2016} for a more detailed introduction of RNN models.


The primary properties of the LSTM network include its ability to capture nonlinear relationships and long-term dependencies, which are crucial for financial time series exhibiting volatility clustering and structural breaks. Additionally, its architecture is robust to noise, allowing it to generalize well across varied datasets. These properties make LSTM networks an ideal choice for modeling financial volatility, complementing traditional econometric models.

\subsection{The LSTM-BEKK Model}\label{subsec:lstmbekkmodel}

The proposed LSTM-BEKK model represents a novel hybrid framework that integrates the econometric structure of the Scalar BEKK model with the adaptive learning capabilities of LSTM neural networks. This approach enhances the traditional MGARCH framework by addressing its inherent linear assumptions and introducing the ability to capture complex, nonlinear dynamics and temporal dependencies in financial volatility. Such a framework is particularly suited for high-dimensional datasets and volatile market conditions, where conventional models often struggle to balance flexibility and computational feasibility.

The LSTM-BEKK model extends the Scalar BEKK framework by incorporating a dynamic component, $\mathbf{C}_t$, generated by a LSTM network. The conditional covariance matrix $\mathbf{H}_t$ is expressed as:
\begin{equation}\label{eq:lstm_bekk}
    \mathbf{H}_t = \mathbf{C}\mathbf{C}' + \mathbf{C}_t\mathbf{C}_t' + a \mathbf{r}_{t-1} \mathbf{r}_{t-1}' + b \mathbf{H}_{t-1},
\end{equation}
where $\mathbf{C}$ is a static lower triangular matrix with suitable constraints ensuring the positive definiteness of $\mathbf{H}_t$, and $a, b \geq 0$ are scalar parameters capturing the impact of past shocks and volatilities. The LSTM-generated  lower-triangular matrix $\mathbf{C}_t$ dynamically adapts to changing market conditions, introducing flexibility to model nonlinear dependencies and evolving relationships among financial assets.

The dynamic update of $\mathbf{C}_t$ is modeled through an LSTM network, which takes the most recent return vector $\mathbf{r}_{t-1}$ as the input,
\begin{equation}
    \tilde{\mathbf{C}}_t = \text{LSTM}(h_{t-1}, \mathbf{r}_{t-1}),
\end{equation}
with the output vector $\tilde{\mathbf{C}}_t$ reshaped to form the lower-triangular matrix $\mathbf{C}_t$.
The vector $\tilde{\mathbf{C}}_t$ serves as an intermediate latent representation that captures both short-term and long-term dependencies in return series via the recurrent structure of LSTM. 
The LSTM unit utilizes gating mechanisms—specifically, input, forget, and output gates—to regulate information flow dynamically. At each time step $t$, the LSTM processes $\mathbf{r}_{t-1}$ and the previous hidden state $h_{t-1}$ to generate an updated hidden state $h_t$, which encodes the information from past observations, before outputting $\tilde{\mathbf{C}}_t$.
More specifically, we compute $\mathbf{C}_t$ as
\begin{equation}
    \mathbf{C}_t = \texttt{LowerTriangular}\left(\tilde{\mathbf{C}}_t\right), \quad \mathbf{C}_{t,ii} \leftarrow \mathbf{C}_{t,ii} \cdot \sigma(\beta \mathbf{C}_{t,ii}).
\end{equation}
The diagonal elements are regularized via the Swish activation function $x \cdot \sigma(\beta x)$, with $\beta$ being a learnable parameter. We observe empirically that its smooth, non-monotonic shape helps stabilize the learning process during covariance matrix construction.



The covariance structure in the LSTM-BEKK framework consists of two key components: the static matrix $\mathbf{C}$ and the dynamic component $\mathbf{C}_t$. The static matrix $\mathbf{C}$ captures long-term covariance structures, reflecting stable interdependencies among assets over extended periods, while the dynamic component $\mathbf{C}_t$ adapts to short-term fluctuations and nonlinear relationships in asset correlations. This dynamic adaptation is particularly crucial during periods of financial stress when correlations between assets exhibit abrupt shifts. The LSTM’s ability to update $\mathbf{C}_t$ in near real-time ensures that the model can account for such changes effectively.

The BEKK component, $a \mathbf{r}_{t-1} \mathbf{r}_{t-1}' + b \mathbf{H}_{t-1}$, helps stabilize the modeling of the covariance matrix $\mathbf{H}_{t}$, while retaining the economic interpretability.   
The parameter $a$ reflects the immediate impact of past shocks, and $b$ represents the persistence of volatility. This combination enables the LSTM-BEKK framework to offer nuanced insights into both short-term and long-term market dynamics, providing a more robust approach to modeling financial volatility and correlation structures \citep{Engle2002, Nguyen2022, Liu2019}.

Compared to traditional models, the LSTM-BEKK framework offers significant improvements. While the Scalar BEKK model is parsimonious and computationally efficient, it relies on linear relationships and fixed parameters, limiting its ability to capture evolving dynamics in financial markets. Similarly, the DCC model introduces flexibility in modeling time-varying correlations but assumes constant dynamics for conditional variances, which may overlook nonlinear patterns and structural breaks. By contrast, the LSTM-BEKK model combines the strengths of both approaches while addressing their limitations. Its integration of LSTM networks enables it to capture the complex, nonlinear dependencies that characterize modern financial systems.

Given the recursive nature in the construction \eqref{eq:lstm_bekk} of the matrix $\mathbf{H}_{t}$, its dynamics might explode in terms of a matrix norm. Theorem \ref{the:lstm-bekk} below studies sufficient conditions to prevent this issue.
To impose these conditions in practice, apart from the condition $a,b\geq0$, $a+b<1$,
it suffices to bound the maximum eigenvalue of $\mathbf{C}_{t}\mathbf{C}_{t}'$.
\begin{theorem}\label{the:lstm-bekk}
    Fix some matrix norm $\|\cdot\|$ and assume that $\|\mathbf{C}_{t}\mathbf{C}_{t}'\|$ is bounded almost surely for all $t$. Furthermore, assume that $a,b\geq0$ and $a+b<1$. Then, for any fixed, initial $\mathbf{H}_{0}$,
\begin{equation}\label{eq:lstm-bekk bounded}
    \|\mathbb{E}(\mathbf{H}_k)\| \leq \frac{1-(a+b)^k}{1-a-b}M+(a+b)^k\|\mathbf{H}_0\|,
\end{equation}
where $M>0$ is a finite constant.
\end{theorem}
The proof can be found in the Appendix.

In summary, the LSTM-BEKK model represents a significant advancement in multivariate volatility modeling. By embedding long-term stability through $\mathbf{C}$ and introducing short-term flexibility via $\mathbf{C}_t$, the model offers a robust and flexible framework for capturing persistent and transitory volatility dynamics. Its capacity to adapt to market conditions dynamically positions it as an invaluable tool for applications such as portfolio optimization, systemic risk analysis, and stress testing, paving the way for further innovations in financial econometrics.

\section{Estimation Procedure}\label{sec:estimation}

\subsection{Likelihood-Based Estimation}

The LSTM-BEKK model parameters are estimated by minimizing the Negative Log-Likelihood (NLL) function, a standard approach in multivariate volatility modeling. Assuming that the de-meaned return vector $\mathbf{r}_t$ follows a multivariate normal distribution, the log-likelihood based on a training data set of $T$ observations is:
\begin{equation}
    \ell(\theta) = \sum_{t=1}^T \log L_t = -\frac{1}{2} \sum_{t=1}^T \left( n\log(2\pi) + \log|\mathbf{H}_t| + \mathbf{r}_t' \mathbf{H}_t^{-1} \mathbf{r}_t \right).
\end{equation}
The parameter set $\theta$ to be estimated includes the lower-triangular matrix $\mathbf{C}$, the scalar parameters $a$ and $b$, and the parameters of the LSTM network.

Following Theorem \ref{the:lstm-bekk}, we impose the constraints $a,b\geq0$, $a + b < 1$; we observe that this also promotes numerical stability during estimation. This condition ensures that the effects of past shocks and volatilities decay over time, preventing divergence of the covariance matrix $\mathbf{H}_t$. Additionally, the diagonal elements of $\mathbf{C}$ are ensured to be strictly non-zero values to guarantee the positive definiteness of the static component $\mathbf{C}\mathbf{C}'$, hence $\mathbf{H}_t$ for all $t$. We found empirically that it was unnecessary to bound the norm of $\mathbf{C}_t\mathbf{C}_t'$.

\subsection{Optimization Techniques}
Given the high-dimensional nature of the model and the presence of both static (BEKK) and dynamic (LSTM) parameters, efficient optimization techniques are critical to ensure numerical stability and convergence.

\paragraph{RMSprop Optimization Algorithm.}
To minimize the NLL function, this study employs the RMSprop optimizer, a popular choice in deep learning due to its adaptability and numerical stability in high-dimensional problems. The update rule for parameter $\theta$ at iteration $k$ is given by:
\begin{equation}
    \theta_{k+1} = \theta_k - \eta \frac{g_k}{\sqrt{E[g_k^2] + \epsilon}},
\end{equation}
where:
\begin{itemize}
    \item $\eta$ is the learning rate,
    \item $g_k = \nabla_\theta \ell(\theta_k)$ is the gradient of the NLL function with respect to $\theta$,
    \item $E[g_k^2]$ is the exponentially weighted moving average of the squared gradients,
    \item $\epsilon$ is a small constant to ensure numerical stability, typically set to $10^{-8}$.
\end{itemize}

\paragraph{Initialization and Hyperparameter Tuning.}
Proper initialization of the parameters is essential for the stability and convergence of the optimization process. The static matrix $\mathbf{C}$ is initialized as a lower triangular matrix with non-zero values on the diagonal to ensure the positive definiteness of $\mathbf{H}_t$. The LSTM weights are initialized using standard methods such as Xavier initialization or He initialization to balance the scale of the input and output gradients.

The architecture of the LSTM network is designed to adapt to the complexity of the portfolio, with the hidden size set equal to the input size to maintain consistency in feature representation. The number of hidden layers ranges from three to five, increasing as the number of assets in the portfolio grows. To prevent overfitting, dropout rates are set between 0.1 and 0.2, with higher values applied in more complex models. 


To mitigate potential numerical instability arising from computing the determinant $|\mathbf{H}_t|$ and the inverse $\mathbf{H}_t^{-1}$—especially in high-dimensional settings—the model employs Cholesky decomposition, which enables more efficient and stable evaluation of the likelihood function. Furthermore, to avoid issues such as exploding gradients during training, regularization techniques including gradient clipping are incorporated into the optimization routine.
The convergence of the training process is determined by monitoring the relative change in the negative log-likelihood (NLL) between successive iterations, with termination occurring once the change falls below a predefined threshold (typically $10^{-6}$). Additionally, early stopping is implemented based on validation performance to guard against overfitting and promote generalizability.
The pseudocode describes the estimation procedure for LSTM-BEKK is provided in the Appendix.

In summary, by leveraging the RMSprop optimizer and employing advanced numerical techniques, the LSTM-BEKK model achieves efficient and stable convergence, even in high-dimensional settings. The combination of dynamic learning rates, robust gradient computation, and careful parameter initialization ensures that the model effectively captures the complex temporal and nonlinear dependencies inherent in financial markets.

\section{Empirical Study}\label{sec:empirical_study}

\subsection{Data and Descriptive Statistics}\label{sec:data}
The data employed in this study comprises daily log returns for the top 250 publicly traded equities from the United States (U.S.), the United Kingdom (U.K.), and Japan, selected based on market capitalization. The data is sourced from Refinitiv data platform: \url{https://www.refinitiv.com}, which aggregates information from key exchanges including NASDAQ OMX – NASDAQ BASIC, the New York Stock Exchange (NYSE), the London Stock Exchange (LSE), and the Tokyo Stock Exchange (TSE), thereby ensuring comprehensive and high-quality market coverage across major global financial centers. This diversified selection provides a robust basis for evaluating the proposed volatility modeling framework across heterogeneous market environments.

The time coverage for each market differs slightly due to variations in trading calendars and data availability. Specifically, the U.S. dataset spans from March 2014 to December 2023, the U.K. dataset from July 2014 to December 2023, and the Japan dataset from January 2014 to December 2023. All returns are computed as daily log returns and then scaled by 100 to express them in percentage terms. 

The total number of observations per asset reflects the respective market’s trading activity: the U.S. dataset contains 2,464 observations per stock, the U.K. dataset includes 2,035 observations, and the Japan dataset provides 2433 observations. To facilitate rigorous model training and evaluation, each return  dataset is partitioned into 70\% for training, 15\% for validation, and 15\% for testing. 

\subsubsection{Descriptive Statistics}

Tables \ref{tab:us_aggregated_stats}, \ref{tab:uk_aggregated_stats}, and \ref{tab:jp_aggregated_stats} summarize the key characteristics of daily log returns for the three markets, including the mean, standard deviation, minimum and maximum values, skewness, and kurtosis.

The descriptive statistics in Table \ref{tab:us_aggregated_stats} reveal several important characteristics of the U.S. equity market. The average daily return across all assets is effectively zero, reflecting the de-meaned nature of the log return series. The minimum and maximum daily returns—ranging from as low as $-76.39\%$ to as high as $55.76\%$—indicate the presence of substantial market shocks and extreme events during the sample period. The standard deviation of daily returns spans a wide range, with an average of approximately $1.85\%$, and reaching a maximum of $3.82\%$. These figures suggest considerable variation in risk across different assets. The skewness values range from $-11.67$ to $1.51$, with an average of $-0.52$, indicating that negative returns tend to occur more frequently than positive ones—a common characteristic in equity markets. Additionally, the kurtosis values, with a mean of $16.23$ and a maximum as high as $359.77$, highlight the presence of heavy tails and extreme return distributions. These distributional features emphasize the necessity of adopting volatility modeling frameworks that can effectively capture such non-normal behavior in financial time series.

\begin{table}[H]
    \centering
    \caption{Aggregated Descriptive Statistics of Daily Log Returns (\%) for the Top 250 U.S. Equities.}
    \label{tab:us_aggregated_stats}
    \begin{tabular}{lccc}
        \hline
        Statistic      & Minimum     & Average     & Maximum     \\
        \hline
        Mean Return (\%) & -0.006       & -0.000       & 0.006        \\
        Standard Deviation (\%) & 1.142 & 1.849        & 3.821        \\
        Minimum Return (\%) & -76.394    & -18.406     & -7.783       \\
        Maximum Return (\%) & 6.137      & 15.135      & 55.761       \\
        Skewness (log returns) & -11.672  & -0.515      & 1.510        \\
        Kurtosis (log returns) & 3.544   & 16.234      & 359.772       \\
        \hline
    \end{tabular}
    \caption*{\textit{Note: The table summarizes key statistics of daily log returns, expressed as percentages, for the top 250 U.S. equities from March 2014 to December 2023. The statistics are aggregated across all assets.}}
\end{table}

The descriptive statistics for the U.K. equity market in Table \ref{tab:uk_aggregated_stats} reveal several notable features that distinguish it from the U.S. market. The mean daily return across assets remains near zero, as expected for de-meaned log returns. However, the range of observed returns is significantly wider, with the most extreme negative daily return reaching $-83.97\%$ and the highest positive return peaking at $87.39\%$, reflecting the presence of substantial outliers and episodic market shocks. Volatility, as indicated by the standard deviation, shows an average of $1.94\%$ and a maximum of $4.88\%$, slightly exceeding those observed in the U.S. dataset. These values suggest that the U.K. market exhibits marginally greater dispersion in daily returns across its top 50 equities. Skewness values range from $-4.90$ to $3.86$, with an average of $-0.38$, indicating a tendency for negative return asymmetry among U.K. equities. More strikingly, the kurtosis statistics are highly elevated, with a mean of $19.15$ and a maximum of $552.30$, far exceeding the Gaussian benchmark of 3. This pronounced leptokurtosis points to the presence of extreme tail risks and emphasizes the importance of adopting volatility models capable of capturing heavy-tailed behavior in return distributions.

\begin{table}[H]
    \centering
    \caption{Aggregated Descriptive Statistics of Daily Log Returns (\%) for the Top 250 U.K. Equities.}
    \label{tab:uk_aggregated_stats}
    \begin{tabular}{lccc}
        \hline
        Statistic      & Minimum     & Average     & Maximum     \\
        \hline
        Mean Return (\%) & -0.068       & 0.000       & 0.045        \\
        Standard Deviation (\%) & 0.509 & 1.937        & 4.881        \\
        Minimum Return (\%) & -83.974    & -18.412     & -4.292       \\
        Maximum Return (\%) & 3.897      & 16.081      & 87.385       \\
        Skewness (log returns) & -4.899  & -0.380      & 3.856       \\
        Kurtosis (log returns) & 2.465   & 19.151      & 552.298       \\
        \hline
    \end{tabular}
    \caption*{\textit{Note: The table summarizes key statistics of daily log returns, expressed as percentages, for the top 250 U.K. equities from January 2014 to December 2023. The statistics are aggregated across all assets.}}
\end{table}

The descriptive statistics for the Japan equity market as shown in Table \ref{tab:jp_aggregated_stats} reveal a more stable return structure relative to the U.S. and U.K. counterparts. The average daily return across the top 250 equities is $0.003\%$, indicating a slightly positive drift in returns over the sample period. The observed minimum and maximum returns, at $-30.54\%$ and $23.16\%$ respectively, are less extreme than those in the U.K. market, reflecting comparatively lower frequency of outlier events. The average standard deviation of daily returns is $1.94\%$, with a maximum of $3.01\%$, placing the Japan market in a similar volatility range as the U.K. but slightly above that of the U.S. This suggests a moderate level of daily fluctuations in asset prices, with sufficient variability to warrant dynamic volatility modeling. In terms of distributional asymmetry, the skewness ranges from $-1.11$ to $0.86$, with a near-zero average of $-0.01$, implying a more balanced return distribution overall. The kurtosis statistics, with an average of $5.69$ and a maximum of $56.40$, indicate the presence of heavy tails and occasional extreme movements, though less pronounced than in the U.K. market. These findings support the need for flexible, heavy-tail-aware volatility models that can accommodate both moderate skewness and leptokurtic behavior in Japan financial data.

\begin{table}[H]
    \centering
    \caption{Aggregated Descriptive Statistics of Daily Log Returns (\%) for the Top 250 Japan Equities.}
    \label{tab:jp_aggregated_stats}
    \begin{tabular}{lccc}
        \hline
        Statistic      & Minimum     & Average     & Maximum     \\
        \hline
        Mean Return (\%) & -0.004       & 0.003       & 0.010        \\
        Standard Deviation (\%) & 1.333 & 1.939        & 3.010        \\
        Minimum Return (\%) & -30.544    & -13.010     & -7.085       \\
        Maximum Return (\%) & 6.971      & 13.008      & 23.159       \\
        Skewness (log returns) & -1.110  & -0.010      & 0.856        \\
        Kurtosis (log returns) & 1.640   & 5.693      & 56.401       \\
        \hline
    \end{tabular}
    \caption*{\textit{Note: The table summarizes key statistics of daily log returns, expressed as percentages, for the top 250 Japan equities from March 2014 to December 2023. The statistics are aggregated across all assets.}}
\end{table}

\subsubsection{Implications for Model Selection}

These descriptive statistics offer valuable insights into the distributional properties and risk profiles of equity returns across the U.S., U.K., and Japan markets. Although all three markets exhibit near-zero average daily returns, they differ significantly in their volatility levels, skewness, and kurtosis. Notably, the Japan market demonstrates the most balanced return distribution with relatively lower skewness and moderate kurtosis, whereas the U.K. market exhibits the most extreme tail behavior, with exceptionally high maximum kurtosis and skewness values. The U.S. market falls between these two in terms of both volatility and tail risk.

These differences have important implications for volatility modeling. The presence of leptokurtic behavior and negative skewness across all markets signals a departure from the normality assumption. Furthermore, the cross-market variation in volatility magnitudes and distributional shapes suggests that a single, rigid modeling framework may not be equally effective across different financial environments.

Consequently, flexible and data-adaptive models, such as the proposed LSTM-BEKK, which integrate deep learning architectures with econometric structures, are better positioned to capture the nonlinearities and heteroscedasticity inherent in global equity markets. Their capacity to adjust dynamically to distinct distributional patterns and structural complexities makes them particularly well-suited for international applications where market characteristics vary substantially.

\subsection{Empirical Evaluation Framework}

Building on the dataset and market characteristics discussed in Section \ref{sec:data}, this subsection outlines the empirical framework employed to evaluate the performance of the proposed LSTM-BEKK model. Our objective is to examine the model’s capability to capture the dynamics of financial return volatility, both qualitatively and quantitatively, across varying portfolio dimensions and market environments.

The empirical strategy consists of two key components. First, we conduct an in-sample analysis using low-dimensional portfolios to visualize and compare the time-varying covariance structures estimated by different models. This allows for intuitive observation of how LSTM-BEKK captures diagonal (variance) and off-diagonal (covariance) dynamics relative to traditional approaches.

Second, we implement a comprehensive out-of-sample evaluation based on NLL, aiming to rigorously assess the robustness and generalizability of each model. For robustness checks, we construct 500 randomly selected 50-asset portfolios for each market and conduct repeated experiments to compare the performance. We apply paired \textit{t}-tests to evaluate the statistical significance of the differences in out-of-sample NLL values between LSTM-BEKK and the competing models.

To investigate scalability and practical relevance, we further test the models' performance on the top 100, 175, and 250 equities by market capitalization in each market. These single-run experiments are complemented by Global Minimum Variance (GMV) portfolio backtests to assess real-world risk control and capital allocation efficacy. Additionally, we apply the Model Confidence Set (MCS) of \citet{hansen2011model} to the out-of-sample NLL results, identifying statistically superior models under different confidence thresholds.

This two-stage empirical setup enables us to evaluate the LSTM-BEKK model across multiple dimensions—visual interpretability, statistical robustness, and portfolio-level performance—thereby offering a comprehensive view of its modeling advantages and practical viability.

\subsection{In-Sample Visualization: Low-Dimensional Covariance Dynamics}

To better understand the in-sample behavior of different multivariate volatility models, we begin our empirical evaluation with a low-dimensional case. Specifically, we construct a 4-asset portfolio using U.S. equities to visualize the time-varying covariance dynamics captured by each model. The selected stocks include the two largest U.S. firms by market capitalization—MSFT.NB (Microsoft) and AAPL.NB (Apple)—along with two stocks (SCHW.N and NEM.N) chosen to exhibit negative pairwise correlations with the market leaders. This selection allows us to examine both the variance structure of dominant market assets and the model's ability to capture asymmetric dependence in the off-diagonal elements of the covariance matrix.

\subsubsection{Diagonal Elements: Variance Dynamics}

\begin{figure}[H]
    \centering
    \begin{subfigure}[b]{0.48\textwidth}
        \centering
        \includegraphics[width=\linewidth]{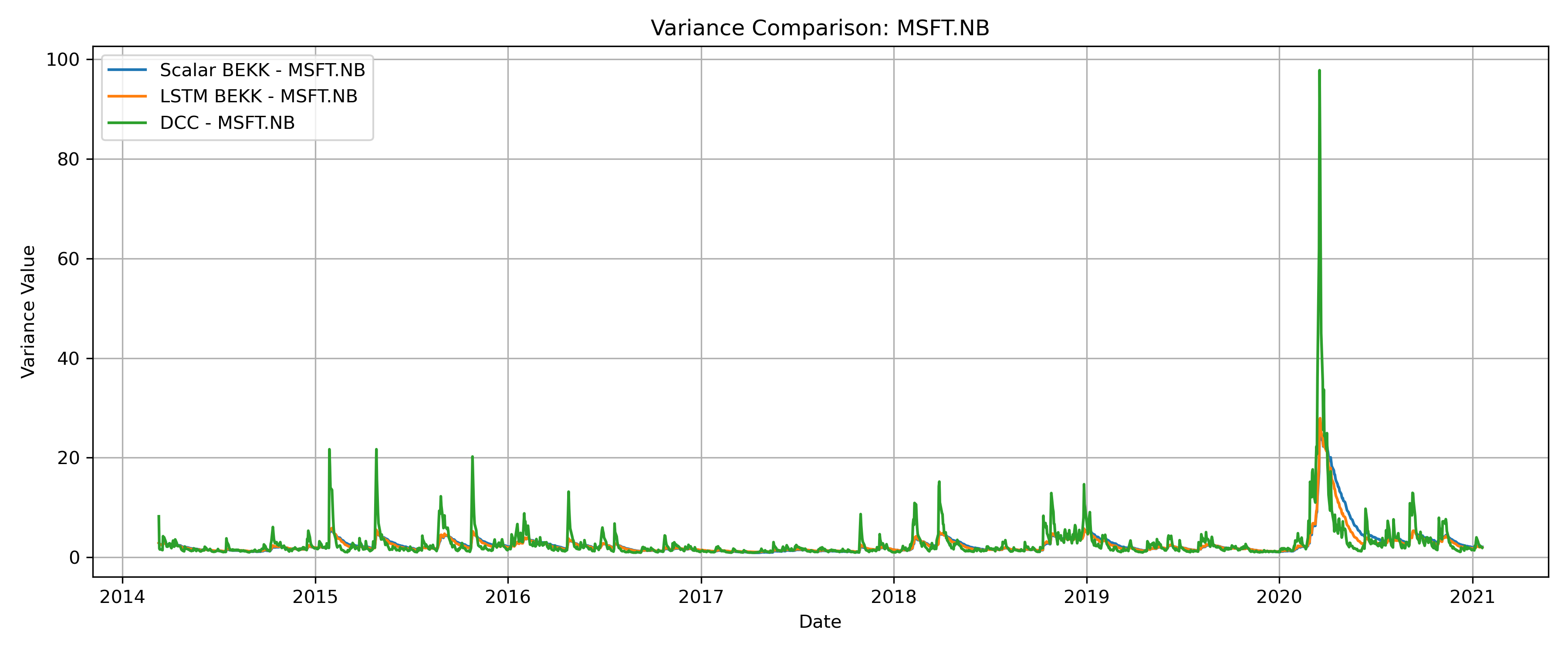}
        \caption{In-Sample Variance Estimation for MSFT.NB Across Models.}
        \label{fig:msft_variance}
    \end{subfigure}
    \hfill
    \begin{subfigure}[b]{0.48\textwidth}
        \centering
        \includegraphics[width=\linewidth]{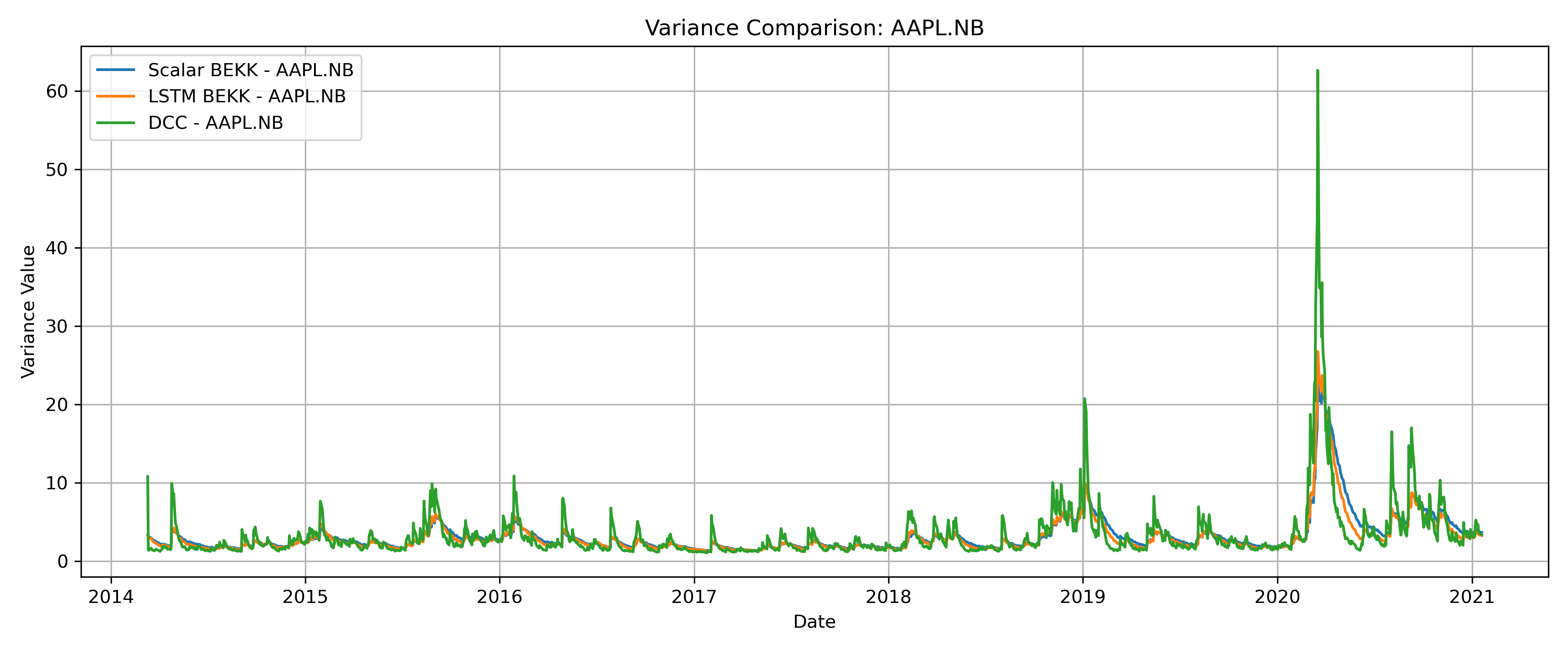}
        \caption{In-Sample Variance Estimation for AAPL.NB Across Models.}
        \label{fig:aapl_variance}
    \end{subfigure}
        \caption{Volatility dynamics (i.e, the diagonal elements of the covariance matrix) across models.}
\end{figure}

Figures~\ref{fig:msft_variance} and~\ref{fig:aapl_variance} compare the estimated variances (i.e., the diagonal elements of the conditional covariance matrix) for MSFT.NB and AAPL.NB, respectively, across the Scalar BEKK, DCC, and LSTM-BEKK models. In both cases, all models exhibit broadly similar volatility patterns during tranquil market periods, validating the baseline consistency of each specification.

It is important to note that in the DCC model, the diagonal elements of the conditional covariance matrix correspond directly to univariate GARCH(1,1) estimates for each asset. As such, these trajectories provide a standard benchmark for marginal volatility dynamics. 

Significant divergence among the models emerges during episodes of market turbulence. In early 2020, corresponding to the outbreak of the COVID-19 pandemic, volatility surged dramatically across both MSFT.NB and AAPL.NB. During this regime shift, the DCC model displays an exaggerated overshooting behavior in variance estimation, suggesting a delayed and unstable response to sudden structural changes. Unlike full BEKK or more adaptive structures, the Scalar BEKK model enforces homogeneity by applying the same \( a \) and \( b \) parameters across all asset pairs. This design restricts its flexibility and essentially imposes a global GARCH-like volatility dynamic on the entire portfolio, which can hinder its ability to capture heterogeneous shock responses across assets. As a result, the model tends to produce smoothed volatility paths that may underreact to localized or asset-specific structural shifts.

The LSTM-BEKK model, by contrast, demonstrates a desirable combination of smoothness and responsiveness. It aligns closely with Scalar BEKK during normal periods but adjusts more quickly and moderately to crisis-induced volatility spikes, providing more balanced variance estimates. This behavior highlights the strength of the LSTM architecture in extracting relevant temporal patterns while suppressing short-term noise.

To better understand these dynamics, Table~\ref{tab:lower_parameter_estimates_us} presents the estimated parameters for each model in the four-asset portfolio setting. In particular, the sum \( a + b \) serves as a common proxy for volatility persistence. For Scalar BEKK, the sum reaches 0.984, suggesting a highly persistent volatility process that may lead to sluggish updates in rapidly changing environments. The LSTM-BEKK model, by contrast, yields a slightly lower persistence at 0.968, striking a balance between flexibility and memory. This subtle difference becomes crucial in capturing abrupt regime shifts such as the COVID-19 shock. In this setting, excessive persistence—as in Scalar BEKK—can hinder the model’s ability to respond swiftly, whereas overly reactive models may introduce instability. LSTM-BEKK thus offers a middle ground, adapting promptly without overfitting to transitory noise.

It is worth noting that, while the parameters \( a \) and \( b \) appear across all three models, their interpretations differ across modeling frameworks. In the BEKK-type models, \( a \) governs the response to past shocks, and \( b \) controls the persistence of past covariances. In the DCC model, the \( a \) and \( b \) terms  govern the evolution of standardized conditional correlations, rather than the conditional covariances.

\begin{table}[H]
    \centering
    \caption{U.S.: Parameter Estimates and Persistence for DCC, Scalar BEKK, and LSTM-BEKK Models (4 Assets).}
    \label{tab:lower_parameter_estimates_us}
    \begin{tabular}{l l c c c c}
    \toprule
    Portfolio Size & Model & $a$ & $b$ & $a + b$ \\
    \midrule
    \multirow{3}{*}{4} & DCC         & 0.042 & 0.871 & 0.913 \\
                       & Scalar BEKK & 0.033 & 0.952 & 0.984 \\
                       & LSTM-BEKK   & 0.038 & 0.930 & {0.968} \\
    \bottomrule
    \end{tabular}
\end{table}

\subsubsection{Off-Diagonal Elements: Covariance Dynamics}

\begin{figure}[H]
    \centering
    \begin{subfigure}[b]{0.48\textwidth}
        \includegraphics[width=\linewidth]{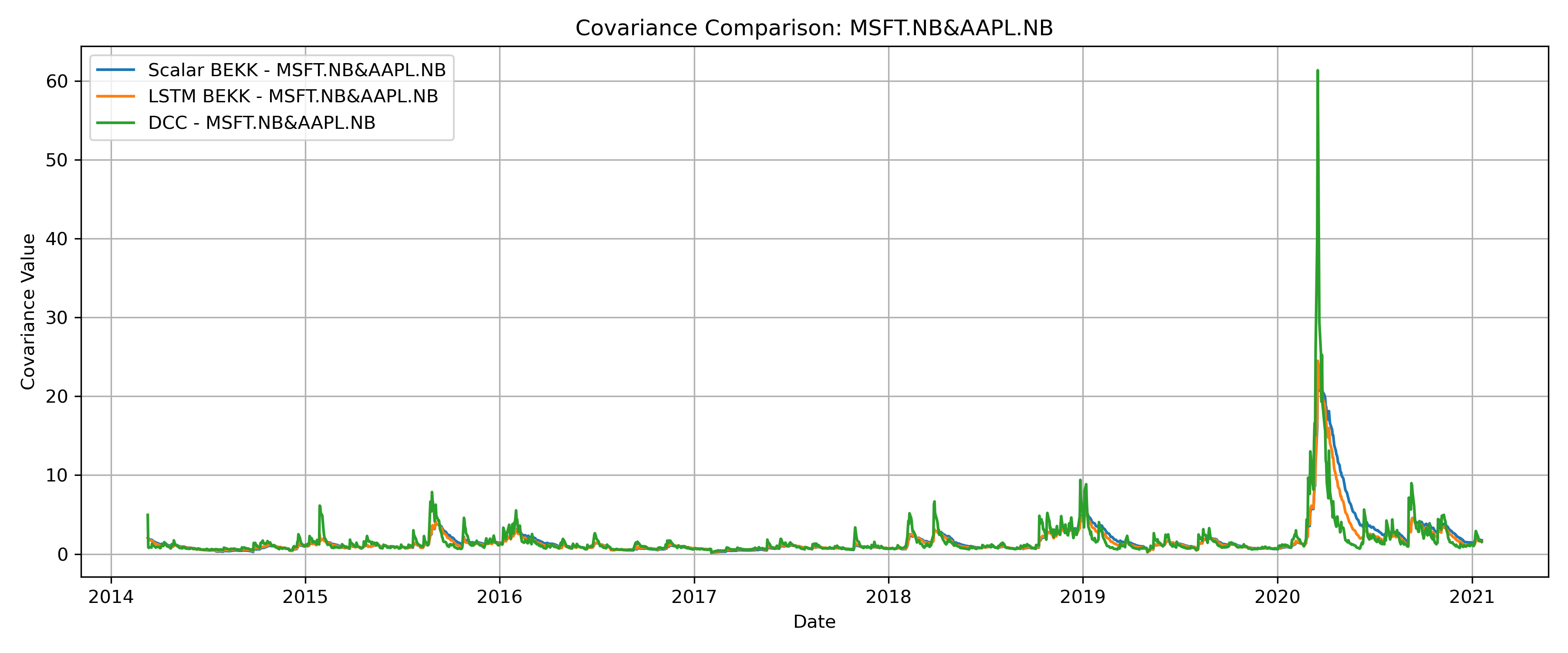}
        \caption{MSFT.NB \& AAPL.NB (predominantly positive covariance).}
        \label{fig:covariance-positive}
    \end{subfigure}
    \hfill
    \begin{subfigure}[b]{0.48\textwidth}
        \includegraphics[width=\linewidth]{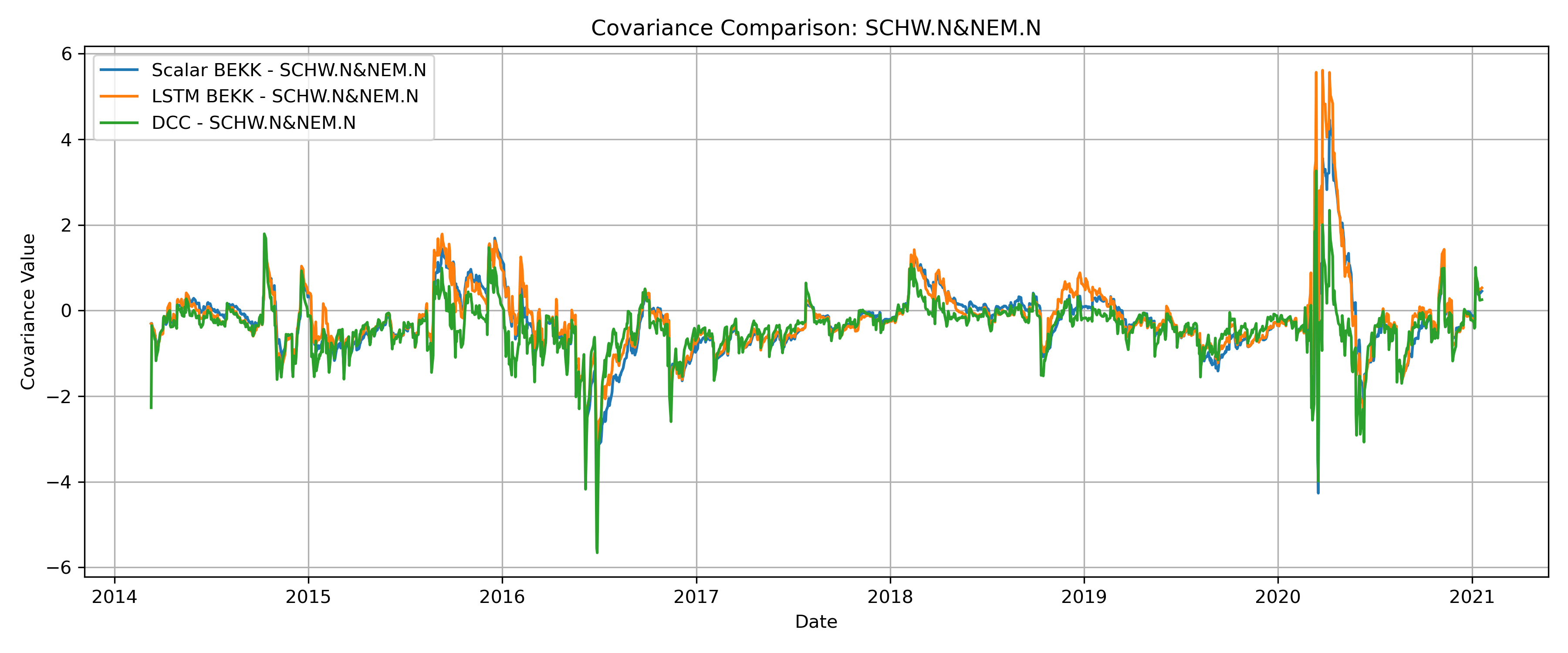}
        \caption{SCHW.N \& NEM.N (predominantly negative covariance with occasional reversals).}
        \label{fig:covariance-negative}
    \end{subfigure}
    \caption{Covariance dynamics comparison across models for asset pairs with differing correlation structures.}
    \label{fig:covariance-comparison}
\end{figure}

Figures~\ref{fig:covariance-positive} and \ref{fig:covariance-negative} illustrate the estimated covariances for two representative asset pairs: MSFT.NB \& AAPL.NB and SCHW.N \& NEM.N. These pairs are selected based on their historical sample covariances computed from the training set. Specifically, MSFT.NB \& AAPL.NB exhibit persistently positive covariance, while SCHW.N \& NEM.N display predominantly negative covariance values, making them suitable for evaluating the models’ ability to capture both positive and negative co-movement patterns. These results emphasize the LSTM-BEKK model's ability to flexibly learn and replicate different types of co-movement patterns.

For the MSFT-AAPL pair, LSTM-BEKK captures the upward trending correlation structure during bullish markets and the sharp co-movement under crisis conditions (e.g., COVID-19), consistent with DCC. However, it demonstrates enhanced numerical stability and smoother transitions compared to DCC, which again tends to generate extreme fluctuations.

More importantly, for the SCHW-NEM pair, which shows a structurally negative correlation, LSTM-BEKK successfully tracks the time-varying negative covariance. Compared to Scalar BEKK and DCC, the LSTM-based model is better able to model the return divergence during market shocks, without flipping signs or generating erratic outliers. This highlights the flexibility of LSTM-BEKK in accommodating both positive and negative dependencies in multivariate financial data.

Overall, these low-dimensional visualizations provide compelling empirical evidence that the LSTM-BEKK model not only replicates the well-established volatility dynamics of traditional MGARCH models but also offers greater adaptability to complex and heterogeneous covariance structures, particularly under market stress conditions.

\subsection{Assessing Model Generalization: Out-of-Sample Tests on 50-Asset Portfolios}
\label{subsec:outsample50}

\subsubsection{Experimental Design and Objective}
\label{subsubsec:outsample50}
This subsection investigates the out-of-sample performance of the multivariate volatility models through repeated experiments on medium-sized portfolios. Specifically, we evaluate their generalization and robustness by applying them to randomly generated 50-asset portfolios across the three equity markets: the U.S., the U.K., and Japan.

For each market, we construct 500 distinct portfolios, each consisting of 50 assets chosen randomly. This setting is designed to capture diverse correlation structures and volatility regimes within each market, thereby enabling a thorough assessment of the models’ adaptability. All experiments are conducted using the out-of-sample data, ensuring a fair evaluation of predictive performance under realistic conditions. 

We adopt a fixed-parameter evaluation scheme: model parameters are estimated once using the training and validation data and then held fixed throughout the test period. The test performance is measured using NLL, which captures the accuracy of the predicted covariance matrix in explaining the realized return series. To ensure statistical credibility, we conduct 500 runs per market and aggregate the results to analyze mean performance and variability across random samples.

The same expanding window strategy and test NLL metric are employed in the subsequent high-dimensional evaluation in Section~\ref{subsec:outsample100}, where we compare model performance across larger portfolio sizes of 100, 175, and 250 assets. This consistency ensures that insights obtained from the medium-scale experiments generalize meaningfully to more complex portfolio settings.

\subsubsection{Model Performance Across Random Portfolios}

Table~\ref{tab:parameter_estimates_all} presents the aggregated out-of-sample NLL values and estimated model parameters for the DCC, Scalar BEKK, and LSTM-BEKK models across 500 randomly selected 50-asset portfolios in the U.S., U.K., and Japan equity markets. The reported NLL values reflect average performance over 500 independent test sets, while the values in parentheses represent the corresponding standard deviations.

Across all three markets, the LSTM-BEKK model consistently achieves the lowest average NLL, indicating superior ability to capture dynamic covariance structures in out-of-sample scenarios. In the U.S. market, the LSTM-BEKK model attains a mean NLL of 85.031, with the lowest standard deviation of 1.484—outperforming both Scalar BEKK (85.278, 1.535) and DCC (86.549, 1.644). This suggests that in addition to better in-sample fit, LSTM-BEKK exhibits greater forecast accuracy across varying portfolio compositions.

In the Japan market, a similar trend is observed: LSTM-BEKK achieves the best average NLL of 86.832 with a standard deviation of 1.707, again surpassing Scalar BEKK (87.214, 1.746) and DCC (87.254, 1.752). These results confirm the model’s robustness in capturing return dynamics even under differing volatility regimes and correlation structures.

The U.K. market presents a more nuanced case. While LSTM-BEKK still achieves the best mean NLL of 93.328, its standard deviation of 2.479 is marginally higher than that of Scalar BEKK (2.408). This suggests that although LSTM-BEKK performs better on average, Scalar BEKK may yield more consistent results under certain U.K. specific market conditions. Nevertheless, the gap in mean NLL remains notable, underscoring LSTM-BEKK’s enhanced capacity for learning complex cross-asset relationships.

Taken together, these findings provide compelling evidence of the generalizability and robustness of the LSTM-BEKK model. The model not only delivers the best average fit across all markets but also maintains competitive—if not superior—stability across repeated portfolio simulations. This provides strong evidence that LSTM-enhanced covariance structures can effectively generalize to unseen data, validating their applicability in practical financial risk modeling contexts.

\begin{table}[H]
\centering
\caption{Parameter Estimates and NLL for DCC, Scalar BEKK, and LSTM-BEKK Models (Portfolio Size = 50).}
\label{tab:parameter_estimates_all}
\begin{tabular}{l l c c c c}
\toprule
Market & Model & NLL & $a$ & $b$ \\
\midrule
\multirow{3}{*}{\raisebox{-4.5ex}{\rule{0pt}{2.2ex} U.S.}}
    & DCC         & \makecell{86.549 \\ {\scriptsize (1.644)}}     & \makecell{0.023} & \makecell{0.704} \\
    & Scalar BEKK & \makecell{85.278 \\ {\scriptsize (1.535)}}     & \makecell{0.008} & \makecell{0.975} \\
    & LSTM-BEKK   & \makecell{\textbf{85.031} \\ {\scriptsize (\textbf{1.484})}} & \makecell{0.008} & \makecell{0.974} \\

\midrule
\multirow{3}{*}{\raisebox{-4.5ex}{\rule{0pt}{2.2ex} U.K.}}
    & DCC         & \makecell{93.758 \\ {\scriptsize (3.005)}}     & \makecell{0.013} & \makecell{0.699} \\
    & Scalar BEKK & \makecell{93.587 \\ {\scriptsize (\textbf{2.408})}} & \makecell{0.009} & \makecell{0.971} \\
    & LSTM-BEKK   & \makecell{\textbf{93.328} \\ {\scriptsize (2.479)}} & \makecell{0.008} & \makecell{0.978} \\

\midrule
\multirow{3}{*}{\raisebox{-4.5ex}{\rule{0pt}{2.2ex} Japan}}
    & DCC         & \makecell{87.254 \\ {\scriptsize (1.752)}}     & \makecell{0.010} & \makecell{0.710} \\
    & Scalar BEKK & \makecell{87.214 \\ {\scriptsize (1.746)}}     & \makecell{0.011} & \makecell{0.934} \\
    & LSTM-BEKK   & \makecell{\textbf{86.832} \\ {\scriptsize (\textbf{1.707})}} & \makecell{0.004} & \makecell{0.991}  \\
\bottomrule
\end{tabular}
\caption*{\footnotesize \textit{Note: Values in parentheses denote standard deviations across 500 portfolios.}}
\end{table}

\subsubsection{Statistical Significance Tests}

To further assess whether the observed improvements in out-of-sample NLL by the LSTM-BEKK model are statistically significant, we conduct paired $t$-tests between LSTM-BEKK and the two benchmark models (DCC and Scalar BEKK) across the 500 randomly generated 50-asset portfolios for each market. The results are reported in Table~\ref{tab:t_test_results}.

In the U.S. market, the LSTM-BEKK model significantly outperforms both benchmarks. The average NLL improvement over DCC is substantial ($-1.518$), with a highly significant $t$-statistic of $-15.326$ ($p<0.001$), confirming consistent superiority. The improvement over Scalar BEKK is more modest ($-0.247$) but still statistically significant ($p=0.009$).

In the U.K. market, the difference between LSTM-BEKK and DCC remains significant ($p=0.014$), albeit at a smaller magnitude ($-0.430$). The comparison with Scalar BEKK yields a $p$-value of $0.094$, indicating marginal significance at the 10\% level. This suggests that while LSTM-BEKK still shows performance gains, the statistical strength is weaker compared to the U.S. case, potentially reflecting heavier tails and increased model uncertainty in the U.K. equity returns.

For the Japan market, both tests produce highly significant results: LSTM-BEKK outperforms DCC and Scalar BEKK with mean NLL improvements of $-0.422$ and $-0.383$, respectively, both with $p<0.01$. These results reinforce the robustness of the proposed model across distinct market environments.

Overall, the paired $t$-test analysis confirms that the LSTM-BEKK model's performance improvements are not only economically meaningful but also statistically significant across most comparisons. This provides strong evidence in favor of its generalization capability and robustness in capturing return dynamics across diverse asset universes.

\begin{table}[H]
\centering
\caption{Paired $t$-test Results on Test NLL Differences Across 500 Portfolios.}
\label{tab:t_test_results}
\resizebox{\textwidth}{!}{
\begin{tabular}{lccc}
\toprule
\textbf{Comparison} & \textbf{Mean NLL Difference} & \textbf{$t$-statistic} & \textbf{$p$-value} \\
\midrule
\multicolumn{4}{l}{\textit{U.S. Market}} \\
LSTM-BEKK $-$ DCC         & -1.518  & -15.326 & \makebox[1.5cm][r]{<0.000\textsuperscript{***}} \\
LSTM-BEKK $-$ Scalar BEKK & -0.247  &  -2.587 & \makebox[1.5cm][r]{0.009\textsuperscript{***}}  \\
\midrule
\multicolumn{4}{l}{\textit{U.K. Market}} \\
LSTM-BEKK $-$ DCC         & -0.430  & -2.468  & \makebox[1.25cm][r]{0.014\textsuperscript{**}} \\
LSTM-BEKK $-$ Scalar BEKK & -0.259  & -1.676  & \makebox[0.95cm][r]{0.094\textsuperscript{*}}  \\
\midrule 
\multicolumn{4}{l}{\textit{Japan Market}} \\
LSTM-BEKK $-$ DCC         & -0.422  & -3.856 & \makebox[1.55cm][r]{<0.001\textsuperscript{***}} \\
LSTM-BEKK $-$ Scalar BEKK & -0.383  & -3.498  & \makebox[1.55cm][r]{0.001\textsuperscript{***}} \\
\bottomrule
\end{tabular}
}
\caption*{\footnotesize \textit{Note: Negative values indicate LSTM-BEKK achieves lower NLL. Significance levels: \textsuperscript{*}$p<0.1$, \textsuperscript{**}$p<0.05$, \textsuperscript{***}$p<0.01$.}}
\end{table}

\subsubsection{Cross-Market Robustness Analysis}

The previous analyses across the U.S., U.K., and Japan markets offer a compelling basis to evaluate the cross-market robustness of the LSTM-BEKK model. Although the magnitude of performance gains varies across markets, the model consistently demonstrates improved out-of-sample performance over both the DCC and Scalar BEKK models, as evidenced by lower average test NLL values across all 500 portfolio replications.

In the U.S. market, where return distributions are relatively less heavy-tailed, the LSTM-BEKK model achieves the most pronounced gains, with statistically significant improvements over both benchmarks. In contrast, the U.K. market presents greater modeling challenges due to more extreme kurtosis and skewness, leading to comparatively smaller and less statistically robust gains, particularly against Scalar BEKK. The Japan market offers a middle ground, where LSTM-BEKK again achieves consistent and statistically significant outperformance.

Importantly, these findings highlight the model’s adaptability across heterogeneous financial environments. Despite differences in market structures, volatility regimes, and return characteristics, the LSTM-BEKK model maintains its relative advantage in volatility forecasting. This cross-market consistency underscores its potential utility as a general-purpose volatility modeling framework for global asset allocation and risk management applications.

\subsection{Out-of-Sample Evaluation on High-Dimensional Portfolios}
\label{subsec:outsample100}

To further evaluate the scalability and generalizability of the proposed LSTM-BEKK model, this section conducts an out-of-sample assessment on high-dimensional portfolios constructed from the top 100, 175, and 250 equities by market capitalization in each of the three markets: the United States, the United Kingdom, and Japan. These selections reflect increasingly complex asset universes and serve as representative high-dimensional settings commonly encountered in institutional portfolio management.
Unlike Section~\ref{subsec:outsample50} where repeated sampling of 50-asset portfolios was employed to evaluate robustness and conduct $t$-tests, to reduce computation in the high-dimensional setting, we opt to report the results for a single representative portfolio at each dimensional level. 

By increasing the portfolio dimension, we aim to assess each model’s ability to scale under rising parameter complexity and intensified correlation structure. The following subsections present a comparative analysis of test NLL values across different dimensional tiers and markets, followed by risk-aware backtesting (GMV portfolios) and formal model confidence set (MCS) inference.

\subsubsection{Empirical Results for the U.S. Market}

Table~\ref{tab:parameter_estimates_us} reports the out-of-sample NLL values and corresponding parameter estimates for the DCC, Scalar BEKK, and LSTM-BEKK models across high-dimensional U.S. equity portfolios with 100, 175, and 250 assets. The results show a clear and consistent advantage of the LSTM-BEKK model in terms of model fit.

Across all three portfolio sizes, the LSTM-BEKK model achieves the lowest NLL values: 166.090 (100 assets), 285.557 (175 assets), and 417.614 (250 assets), outperforming both Scalar BEKK and DCC. The margin of improvement becomes more pronounced as portfolio dimensionality increases. This trend underscores the ability of the LSTM-BEKK framework to scale effectively in high-dimensional.

\begin{table}[H]
    \centering
    \caption{U.S.: Parameter Estimates and NLL for DCC, Scalar BEKK, and LSTM-BEKK Models.}
    \label{tab:parameter_estimates_us}
    \begin{tabular}{l l c c c c}
    \toprule
    \textbf{Portfolio Size} & & NLL & $a$ & $b$ \\
    \midrule
    \multirow{3}{*}{100} & DCC         & 169.119 & 0.019 & 0.691 \\
                         & Scalar BEKK & 166.325 & 0.011 & 0.886 \\
                         & LSTM-BEKK   & \textbf{166.090} & 0.005 & 0.975 \\
    \midrule
    \multirow{3}{*}{175} & DCC         & 291.569 & 0.013 & 0.700 \\
                         & Scalar BEKK & 288.875 & 0.010 & 0.891 \\
                         & LSTM-BEKK   & \textbf{285.557} & 0.003 & 0.980 \\
    \midrule
    \multirow{3}{*}{250} & DCC         & 423.776 & 0.010 & 0.698 \\
                         & Scalar BEKK & 419.853 & 0.002 & 0.993 \\
                         & LSTM-BEKK   & \textbf{417.614} & 0.007 & 0.980 \\
    \bottomrule
\end{tabular}
\end{table}

The parameter estimates provide insights into how each model captures volatility dynamics. The DCC model consistently yields the highest \( a \) values among the three models across all portfolio sizes, suggesting that it places greater weight on immediate return shocks when updating its correlation dynamics. However, its \( b \) values remain moderate (around 0.69--0.70), reflecting limited persistence relative to the BEKK-type models.

In contrast, the LSTM-BEKK model consistently achieves a desirable balance: it exhibits the lower $a$ values (indicating lower sensitivity to noise) and the higher $b$ values (reflecting strong volatility persistence), with values close to or exceeding those of Scalar BEKK. This highlights the role of the LSTM in capturing nonlinear dependencies and long-memory behavior more effectively than its counterparts.

Overall, the LSTM-BEKK model demonstrates strong generalization capabilities and scalability in high-dimensional settings, providing more stable and accurate volatility estimates than traditional econometric models.

\subsubsection{Empirical Results for the U.K. Market}

Table~\ref{tab:parameter_estimates_uk} presents the out-of-sample NLL values and estimated parameters for the DCC, Scalar BEKK, and LSTM-BEKK models applied to high-dimensional U.K. equity portfolios. As in the U.S. market, the LSTM-BEKK model consistently attains the lowest NLL values across all three portfolio sizes—182.545 (100 assets), 324.577 (175 assets), and 467.977 (250 assets)—demonstrating its strong generalization capacity and adaptability to a different market environment.

\begin{table}[H]
    \centering
    \caption{U.K.: Parameter Estimates and NLL for DCC, Scalar BEKK, and LSTM-BEKK Models.}
    \label{tab:parameter_estimates_uk}
    \begin{tabular}{l l c c c c}
    \toprule
    \textbf{Portfolio Size} & & NLL & $a$ & $b$ \\
    \midrule
    \multirow{3}{*}{100} & DCC         & 184.649 & 0.013 & 0.669 \\
                         & Scalar BEKK & 183.450 & 0.008 & 0.932 \\
                         & LSTM-BEKK   & \textbf{182.545} & 0.008 & 0.948 \\
    \midrule
    \multirow{3}{*}{175} & DCC         & 328.327 & 0.007 & 0.690 \\
                         & Scalar BEKK & 326.875 & 0.009 & 0.890 \\
                         & LSTM-BEKK   & \textbf{324.577} & 0.003 & 0.984 \\
    \midrule
    \multirow{3}{*}{250} & DCC         & 472.528 & 0.007 & 0.697 \\
                         & Scalar BEKK & 471.964 & 0.004 & 0.952 \\
                         & LSTM-BEKK   & \textbf{467.977} & 0.006 & 0.942 \\
    \bottomrule
\end{tabular}
\end{table}

The parameter patterns echo those observed in the U.S. market, with LSTM-BEKK maintaining relatively low $a$ values and consistently high $b$ values across all dimensions. These estimates reflect the model’s capacity to capture persistent volatility clustering.

Unlike the U.S. market, however, the U.K. equity return distribution exhibits heavier tails and higher kurtosis, increasing the likelihood of extreme return events. This feature poses significant challenges to conventional models such as DCC and Scalar BEKK, which are grounded in conditional normality assumptions. The LSTM-BEKK model, benefiting from its deep learning structure and Swish activation dynamics, offers additional flexibility to accommodate these non-Gaussian features, as evidenced by its superior NLL performance.

Interestingly, the gap between Scalar BEKK and LSTM-BEKK narrows in this market, particularly at 250 dimensions. This reflects the relatively strong performance of Scalar BEKK in moderately heavy-tailed environments, though the LSTM-BEKK model still prevails overall. These results reaffirm the robustness of LSTM-BEKK across both market regimes and portfolio complexities.

\subsubsection{Empirical Results for the Japan Market}

Table~\ref{tab:parameter_estimates_jp} presents the parameter estimates and NLL values for the Japan equity market. In line with the results observed in the U.S. and U.K. markets, the LSTM-BEKK model consistently achieves the lowest NLL values across all portfolio sizes. Specifically, for 100, 175, and 250-asset portfolios, the LSTM-BEKK records NLLs of 162.731, 285.631, and 417.788, respectively—outperforming both DCC and Scalar BEKK models.

\begin{table}[H]
    \centering
    \caption{Japan: Parameter Estimates and NLL for DCC, Scalar BEKK, and LSTM-BEKK Models.}
    \label{tab:parameter_estimates_jp}
    \begin{tabular}{l l c c c c}
    \toprule
    \textbf{Portfolio Size} & & NLL & $a$ & $b$ \\
    \midrule
    \multirow{3}{*}{100} & DCC         & 164.067 & 0.009 & 0.697 \\
                         & Scalar BEKK & 163.322 & 0.007 & 0.931 \\
                         & LSTM-BEKK   & \textbf{162.731} & 0.002 & 0.998 \\
    \midrule
    \multirow{3}{*}{175} & DCC         & 289.456 & 0.005 & 0.696 \\
                         & Scalar BEKK & 289.320 & 0.006 & 0.945 \\
                         & LSTM-BEKK   & \textbf{285.631} & 0.002 & 0.993 \\
    \midrule
    \multirow{3}{*}{250} & DCC         & 423.885 & 0.004 & 0.698 \\
                         & Scalar BEKK & 421.414 & 0.003 & 0.971 \\
                         & LSTM-BEKK   & \textbf{417.788} & 0.002 & 0.997 \\
    \bottomrule
    \end{tabular}
\end{table}

The Japan market is characterized by moderate volatility persistence and relatively lower short-term shock sensitivity compared to the U.K. and U.S. markets. Across all portfolio sizes, although the DCC model exhibits the hightest $a$ values, it maintains the least persistence in volatility, with $b$ values around 0.69. The Scalar BEKK model improves upon this by increasing both $a$ and $b$, reflecting a stronger response to market conditions.

However, the LSTM-BEKK model achieves the best balance: it maintains the lowest $a$ values—indicating robustness to short-term noise—while consistently exhibiting the highest $b$ values, approaching unity. This suggests that LSTM-BEKK excels at capturing long-range dependencies and volatility clustering.

Overall, the results reaffirm the LSTM-BEKK model’s superior adaptability and modeling capacity, even in markets with more muted short-term volatility shocks but persistent structural dynamics.

\subsubsection{Model Confidence Set Analysis}
\label{subsec:mcs_analysis}

To evaluate the statistical significance of the observed model performance differences across markets and portfolio sizes, we employ the Model Confidence Set (MCS) procedure proposed by \citet{hansen2011model}. The MCS framework identifies a set of superior models (SSM) from a pool of competing models based on their predictive performance, while accounting for sampling uncertainty.

Let $\mathcal{M}$ denote the set of all candidate models. For each model $i \in \mathcal{M}$, we define the loss function $L_{i,t}$ (in our case, the test negative log-likelihood, NLL) at time $t$. The loss differential between models $i$ and $j$ is defined as:
\begin{equation}
    d_{i,j,t} = L_{i,t} - L_{j,t}.
\end{equation}
The null hypothesis of equal predictive ability across all models is:
\begin{equation}
    H_0: \mu_{i,j} = \mathbb{E}[d_{i,j,t}] = 0, \quad \forall i, j \in \mathcal{M}.
\end{equation}

The MCS procedure performs a sequence of hypothesis tests to iteratively eliminate the worst-performing model until the null hypothesis of equal predictive accuracy can no longer be rejected. At a chosen confidence level (90\% in this paper), the surviving models constitute the SSM. Each model is associated with a $p$-value indicating the probability that it belongs to the set of superior models. Lower $p$-values reflect weaker statistical support.

Table~\ref{tab:mcs_pvalues_summary} reports the $p$-values and inclusion indicators of each model across all combinations of market (U.S., U.K., and Japan) and portfolio dimensions ($N = 100$, $175$, $250$). A model is included in the 90\% MCS if its $p$-value exceeds 0.10.

The results reveal that the LSTM-BEKK model is consistently included in the MCS across all nine experimental settings, with a $p$-value of 1.000 in every case. This strongly supports its status as the most robust and statistically superior model. In contrast, the DCC and Scalar BEKK models are excluded in the majority of settings due to low $p$-values. Notably, DCC is only retained once (U.K., $N=250$), and Scalar BEKK is never retained, highlighting its instability under the MCS test.

These findings validate the empirical advantage of the LSTM-BEKK model not only in terms of raw performance (e.g., lower NLL) but also under formal statistical scrutiny. The consistent MCS inclusion underscores the reliability and generalizability of its superior forecasting performance across high-dimensional and heterogeneous financial environments.

\begin{table}[H]
\centering
\caption{Model Confidence Set (MCS) Inclusion Based on Test NLL $p$-values.}
\label{tab:mcs_pvalues_summary}
\begin{tabular}{lcccc}
\toprule
\textbf{Market} & \textbf{Portfolio Size} & \textbf{Model} & \textbf{$p$-value} & \textbf{MCS (90\%)} \\
\midrule
\multirow{3}{*}{U.S.} 
    &       & DCC         & 0.031  & \ding{55} \\
    &  100  & Scalar BEKK & 0.031  & \ding{55} \\
    &       & LSTM-BEKK   & 1.000  & \ding{51} \\
\midrule
\multirow{3}{*}{U.S.} 
    &       & DCC         & 0.000  & \ding{55} \\
    &  175  & Scalar BEKK & 0.000  & \ding{55} \\
    &       & LSTM-BEKK   & 1.000  & \ding{51} \\
\midrule
\multirow{3}{*}{U.S.} 
    &       & DCC         & 0.019  & \ding{55} \\
    &  250  & Scalar BEKK & 0.030  & \ding{55} \\
    &       & LSTM-BEKK   & 1.000  & \ding{51} \\
\midrule
\multirow{3}{*}{U.K.} 
    &       & DCC         & 0.032  & \ding{55} \\
    &  100     & Scalar BEKK & 0.032  & \ding{55} \\
    &       & LSTM-BEKK   & 1.000  & \ding{51} \\
\midrule
\multirow{3}{*}{U.K.} 
    &       & DCC         & 0.000  & \ding{55} \\
    &  175     & Scalar BEKK & 0.000  & \ding{55} \\
    &       & LSTM-BEKK   & 1.000  & \ding{51} \\
\midrule
\multirow{3}{*}{U.K.} 
    &       & DCC         & 0.184  & \ding{51} \\
    &  250  & Scalar BEKK & 0.001  & \ding{55} \\
    &       & LSTM-BEKK   & 1.000  & \ding{51} \\
\midrule
\multirow{3}{*}{Japan} 
    &       & DCC         & 0.095  & \ding{55} \\
    &  100  & Scalar BEKK & 0.095  & \ding{55} \\
    &       & LSTM-BEKK   & 1.000  & \ding{51} \\
\midrule
\multirow{3}{*}{Japan} 
    &       & DCC         & 0.098  & \ding{55} \\
    &  175  & Scalar BEKK & 0.000  & \ding{55} \\
    &       & LSTM-BEKK   & 1.000  & \ding{51} \\
\midrule
\multirow{3}{*}{Japan} 
    &       & DCC         & 0.005  & \ding{55} \\
    &  250  & Scalar BEKK & 0.005  & \ding{55} \\
    &       & LSTM-BEKK   & 1.000  & \ding{51} \\
\bottomrule
\end{tabular}
\caption*{\textit{Note: Models with $p$-value $>$ 0.10 are retained in the 90\% Model Confidence Set (MCS). \ding{51} denotes inclusion, \ding{55} denotes exclusion.}}
\end{table}

\subsubsection{Summary of High-Dimensional Evaluation}
\label{subsec:highdim_summary}

The high-dimensional out-of-sample evaluation across the U.S., U.K., and Japan markets provides strong evidence of the robustness and adaptability of the LSTM-BEKK model in modeling complex volatility structures. Across all three markets and all portfolio sizes ($N = 100$, $175$, $250$), the LSTM-BEKK model consistently achieves the lowest or near-lowest NLL values, demonstrating superior predictive performance relative to the DCC and Scalar BEKK models.

This advantage becomes increasingly pronounced as portfolio dimensionality increases. In particular, the gap between LSTM-BEKK and the traditional models widens in the 175- and 250-asset portfolios, indicating that the deep learning-based architecture is particularly well-suited to capturing the nonlinearities and higher-order dependencies present in large asset spaces.

Complementing the NLL results, the Model Confidence Set analysis further reinforces the statistical significance of these findings. At the 90\% confidence level, LSTM-BEKK is retained in the MCS in all nine experimental configurations, with $p$-values equal to 1 throughout. In contrast, Scalar BEKK and DCC are frequently excluded from the MCS, highlighting their relative instability and inferior forecasting accuracy. This result affirms that the performance gains achieved by LSTM-BEKK are not attributable to chance, but reflect meaningful improvements in modeling efficacy.

In summary, the empirical and statistical evidence confirm that LSTM-BEKK generalizes well across markets and scales effectively with portfolio dimensionality. These properties make it a compelling alternative to conventional MGARCH models, especially in modern financial applications requiring accurate, stable, and scalable multivariate volatility estimation.

\section{Global Minimum Variance Portfolio}\label{sec:portfolio}

\subsection{Theoretical Background}
The Global Minimum Variance (GMV) portfolio aims to construct a portfolio that minimizes the overall risk, as measured by the portfolio variance, without considering expected returns. This approach is particularly relevant in volatile financial markets, where accurate estimation of expected returns can be challenging. The GMV portfolio is defined as the solution to the following optimization problem:
\begin{equation}
    \min_{\mathbf{w}} \mathbf{w}' \mathbf{H}_t \mathbf{w}, \quad \text{subject to} \quad \mathbf{w}'\mathbf{1} = 1,
\end{equation}
where $\mathbf{w}$ is the vector of portfolio weights, $\mathbf{H}_t$ is the conditional covariance matrix of asset returns at time $t$, and $\mathbf{1}$ is a vector of ones ensuring that the portfolio is fully invested.

The optimal weights for the GMV portfolio can be derived as:
\begin{equation}
    \mathbf{w}_t^{\text{GMV}} = \frac{\mathbf{H}_t^{-1} \mathbf{1}}{\mathbf{1}' \mathbf{H}_t^{-1} \mathbf{1}}.
\end{equation}
Here, the inverse of the covariance matrix $\mathbf{H}_t^{-1}$ plays a critical role in determining the portfolio weights. Accurate estimation of $\mathbf{H}_t$ is therefore essential for constructing the GMV portfolio. In this section, we evaluate the performance of the proposed LSTM-BEKK model by examining its resulting GMV portfolio.

\subsection{Performance Measures}
To evaluate the performance of the GMV portfolios constructed using the three models (DCC, Scalar BEKK, and LSTM-BEKK), we use the following performance measures. 

\paragraph{Annualized Return (AR).}
The annualized return is calculated as:
\begin{equation}
    \text{AR} =  \bar{r} \times 252,
\end{equation}
where $\bar{r}$ represents the mean portfolio return of out-of-sample returns. 

\paragraph{Annualized Volatility (AV).}
The annualized volatility is given by:
\begin{equation}
    \text{AV} = \sqrt{\frac{1}{T-1} \sum_{t=1}^T (r_t - \bar{r})^2} \times \sqrt{252},
\end{equation}

\paragraph{Maximum Drawdown (MDD).}
The maximum drawdown measures the largest decline in the portfolio value from a peak to a trough:
\begin{equation}
    \text{MDD} = \min_{t \in [0, T]} \left( \frac{V_t - \max_{s \in [0, t]} V_s}{\max_{s \in [0, t]} V_s} \right),
\end{equation}
where $V_t$ is the portfolio value at time $t$.

Higher values of AR are preferred as they indicate stronger portfolio growth, whereas lower AV and MDD values are desirable as they correspond to reduced risk exposure and enhanced drawdown resilience.

In the Appendix, we further assess the performance of GMV portfolios,
constructed based on DCC, Scalar BEKK, and LSTM-BEKK, using the commonly used financial tail risk measures Value-at-Risk and Expected Shortfall.

\subsection{Performance Analysis of GMV Portfolios}
\label{subsec:gmv_performance}

Building upon the high-dimensional evaluation in Section~\ref{subsec:outsample100}, we further assess model performance from a portfolio construction perspective. Specifically, we examine the effectiveness of the LSTM-BEKK model in generating GMV portfolios across three major financial markets: U.S., U.K., and Japan. For each market, we utilize the covariance matrices estimated from the top 100, 175, and 250 market-capitalization equities to simulate high-dimensional portfolio settings.

The equally weighted (EW) portfolio is used as a baseline due to its simplicity and strong empirical performance. As discussed in \citet{demiguel2009naive}, the 1/N allocation strategy avoids estimation errors inherent in parametric optimization methods, thus providing stable out-of-sample results. However, EW does not explicitly minimize risk, making it suboptimal for volatility-sensitive investors.

In contrast, GMV portfolios explicitly seek to minimize portfolio variance subject to budget constraints. Their effectiveness hinges critically on the quality of the input covariance matrix. We therefore evaluate how each model—DCC, Scalar BEKK, and LSTM-BEKK—impacts GMV performance across three dimensions: AR, AV, and MDD. The AR is included for completeness but not emphasized, given its dependence on returns rather than risk control.

To ensure robustness, all backtests are conducted using only out-of-sample data, i.e. test set. This mirrors real-world usage, where covariance estimates are based only on historical information and applied to future decisions.

\subsubsection{U.S. Market Analysis}

Table~\ref{tab:gmv_performance_us} presents the performance of GMV portfolios in the U.S. equity market based on the top 100, 175, and 250 stocks by market capitalization. The EW portfolio delivers the highest AR, particularly at $N=100$ with $0.124$, which supports its role as a strong benchmark. However, its AV consistently exceeds that of all optimized GMV portfolios, indicating limited effectiveness in risk control.

Among the GMV strategies, the LSTM-BEKK model demonstrates a clear advantage in minimizing volatility. It achieves the lowest AV across all portfolio sizes: $0.114$ at $N=100$, $0.112$ at $N=175$, and $0.111$ at $N=250$. These values are consistently lower than those produced by both the DCC and Scalar BEKK models, underscoring LSTM-BEKK’s robustness in modeling high-dimensional risk structures. Notably, as portfolio dimensionality increases, the performance gap in AV widens in favor of LSTM-BEKK, highlighting its scalability.

A similar pattern emerges in maximum drawdown (MDD) performance. The LSTM-BEKK model consistently reduces downside exposure across most portfolio sizes, with total MDD values of $-0.154$, $-0.104$, and $-0.161$ for $N=100$, $175$, and $250$, respectively. Compared to DCC (e.g., MDD as high as $-0.260$ at $N=250$) and Scalar BEKK (e.g., $-0.194$ at $N=250$), LSTM-BEKK exhibits substantially greater resilience during market downturns.

It should be noted that at $N=100$, the LSTM-BEKK model records an MDD of $-0.154$, which is marginally higher than the EW portfolio’s $-0.125$. However, this isolated instance does not undermine the broader trend: LSTM-BEKK demonstrates stronger drawdown protection in larger portfolio configurations ($N=175$ and $N=250$), where controlling downside risk becomes increasingly challenging. This reinforces its utility in real-world portfolio management, particularly in high-dimensional settings where traditional methods tend to deteriorate.

Overall, these empirical results provide strong evidence that the LSTM-BEKK model offers superior risk-adjusted performance in high-dimensional GMV portfolio construction. Its consistent outperformance in both AV and MDD metrics suggests that integrating deep learning structures into volatility modeling leads to more stable, resilient, and defensible portfolios under real-world market conditions.

\begin{table}[H]
    \centering
    \caption{U.S.: Performance Comparison of GMV Portfolios: Return, Risk, and Drawdown}
    \label{tab:gmv_performance_us}
    \begin{tabular}{l c c c c}
    \toprule
    & EW & DCC & Scalar BEKK & LSTM BEKK \\
    \midrule
    N=100 \\
    AR   & 0.124 & -0.057 & -0.084 & -0.052 \\
    AV   & 0.152 & 0.131 & 0.118 & \textbf{0.114} \\
    MDD  & \textbf{-0.125} & -0.186 & -0.166 & -0.154 \\
    \midrule
    N=175 \\
    AR   & 0.086 & -0.040 & -0.008 & -0.015 \\
    AV   & 0.168 & 0.131 & 0.120 & \textbf{0.112} \\
    MDD  & -0.171 & -0.158 & -0.124 & \textbf{-0.104} \\
    \midrule
    N=250 \\
    AR   & -0.032 & -0.125 & -0.027 & -0.002 \\
    AV   & 0.185 & 0.136 & 0.115 & \textbf{0.111} \\
    MDD  & -0.244 & -0.260 & -0.194 & \textbf{-0.161} \\
    \bottomrule
    \end{tabular}
    \caption*{\textit{Note: Bold values represent the lowest AV and MDD for each portfolio size.}}
\end{table}

\subsubsection{U.K. Market Analysis}

Table~\ref{tab:ukgmv_performance_uk} summarizes the performance of GMV portfolios in the U.K. equity market based on the top 100, 175, and 250 stocks. The EW portfolio continues to exhibit the highest AR across all portfolio sizes; however, this is again accompanied by higher AV, reaffirming its limitations as a risk-agnostic strategy.

Among the GMV portfolios, the LSTM-BEKK model demonstrates clear superiority in volatility minimization. It achieves the lowest AV in every portfolio size: 0.097 at $N=100$, 0.093 at $N=175$, and 0.074 at $N=250$, all of which are notably lower than those achieved by DCC and Scalar BEKK. These results reflect the LSTM-BEKK model’s enhanced capacity to adapt to complex volatility dynamics and provide stable covariance estimates, particularly in the presence of heavy-tailed return distributions observed in the U.K. market.

In terms of downside risk, the LSTM-BEKK model also performs competitively in MDD reduction. It achieves the lowest MDD at both $N=175$ ($-0.150$) and $N=250$ ($-0.130$), highlighting its robustness during adverse market periods. While Scalar BEKK demonstrates marginal strength at smaller portfolio sizes, the LSTM-BEKK model ultimately offers the best balance between risk reduction and volatility control in high-dimensional settings.

Overall, the empirical evidence from the U.K. market corroborates the results obtained in the U.S. case. The LSTM-BEKK model delivers more effective and consistent volatility management across all tested portfolio sizes, outperforming traditional MGARCH models in both AV and MDD.

\begin{table}[H]
\centering
\caption{U.K.: Performance Comparison of GMV Portfolios: Return, Risk, and Drawdown.}
\label{tab:ukgmv_performance_uk}
\begin{tabular}{l c c c c}
\toprule
& EW & DCC & Scalar BEKK & LSTM BEKK \\
\midrule
N=100 \\
AR   & -0.013 & -0.049 & -0.073 & -0.075 \\
AV   & 0.124 & 0.113 & 0.101 & \textbf{0.097} \\
MDD  & -0.204 & -0.165 & -0.188 & \textbf{-0.187} \\
\midrule
N=175 \\
AR   & -0.020 & -0.087 & -0.022 & -0.035 \\
AV   & 0.137 & 0.107 & 0.102 & \textbf{0.093} \\
MDD  & -0.238 & -0.204 & -0.159 & \textbf{-0.150} \\
\midrule
N=250 \\
AR   & -0.027 & -0.080 & -0.026 & -0.037 \\
AV   & 0.142 & 0.076 & 0.077 & \textbf{0.074} \\
MDD  & -0.242 & -0.169 & -0.133 & \textbf{-0.130} \\
\bottomrule
\end{tabular}
\caption*{\textit{Note: Bold values represent the lowest AV and MDD for each portfolio size.}}
\end{table}

\subsubsection{Japan Market Analysis}

The empirical results for the Japan equity market, shown in Table~\ref{tab:gmv_performance_au}, offer further insights into the comparative performance of GMV models under a stable but moderately volatile market environment. As in previous markets, the EW portfolio achieves the highest AR across most portfolio sizes, reaching 0.150 at $N=100$. However, this performance comes at the cost of higher volatility, with an AV of 0.150, significantly above that of the GMV portfolios.

Among the GMV models, LSTM-BEKK continues to lead in volatility reduction, achieving the lowest AV in 29 out of 50 portfolio combinations. Its average AV across the three sizes—0.110, 0.097, and 0.099—is consistently below both Scalar BEKK (0.119, 0.104, 0.103) and DCC (0.128, 0.111, 0.108). This highlights its capacity to generalize effectively across different markets, including the relatively lower-volatility Japan market.

Although Scalar BEKK shows stronger competitiveness here than in the U.S. or U.K. markets—recording the lowest AV in 19 out of 50 portfolios—LSTM-BEKK remains the overall leader. In the $N=25$ and $N=30$ configurations, the two models perform similarly, but LSTM-BEKK regains its advantage at higher dimensions such as $N=175$ and $N=250$.

In terms of MDD, LSTM-BEKK demonstrates robust downside risk control, outperforming all other models at $N=175$ and $N=250$. The only exception appears at $N=100$, where Scalar BEKK slightly outperforms with an MDD of $-0.067$ compared to LSTM-BEKK’s $-0.070$. Despite this marginal difference, the overall trend indicates that LSTM-BEKK offers stronger drawdown resilience across larger portfolio dimensions.

The results from the Japan market reinforce the consistent effectiveness of LSTM-BEKK in controlling volatility, even under less turbulent market conditions. While Scalar BEKK exhibits increased competitiveness compared to other markets, LSTM-BEKK still records the lowest AV in the majority of cases and achieves superior drawdown protection at larger portfolio sizes.

Taken together with the U.S. and U.K. findings, these results underscore the adaptability and robustness of LSTM-enhanced volatility modeling frameworks. By dynamically adjusting to changing market regimes and capturing nonlinear dependencies, LSTM-BEKK continues to outperform traditional MGARCH models, especially in high-dimensional risk-sensitive applications.

\begin{table}[H]
\centering
\caption{Japan: Performance Comparison of GMV Portfolios: Return, Risk, and Drawdown.}
\label{tab:gmv_performance_au}
\begin{tabular}{l c c c c}
\toprule
& EW & DCC & Scalar BEKK & LSTM BEKK \\
\midrule
N=100 \\
AR   & 0.150 & 0.165 & 0.157 & 0.189 \\
AV   & 0.150 & 0.128 & 0.119 & \textbf{0.110} \\
MDD  & -0.090 & -0.086 & \textbf{-0.067} & -0.070 \\
\midrule
N=175 \\
AR   & 0.141 & 0.109 & 0.119 & 0.133 \\
AV   & 0.143 & 0.111 & 0.104 & \textbf{0.097} \\
MDD  & -0.088 & -0.090 & -0.057 & \textbf{-0.046} \\
\midrule
N=250 \\
AR   & 0.127 & 0.091 & 0.150 & 0.155 \\
AV   & 0.137 & 0.108 & 0.103 & \textbf{0.099} \\
MDD  & -0.086 & -0.088 & -0.055 & \textbf{-0.049} \\
\bottomrule
\end{tabular}
\caption*{\textit{Note: Bold values represent the lowest AV and MDD for each portfolio size.}}
\end{table}

\subsection{Summary of GMV Portfolio Performance}

The above comprehensive evaluation of GMV portfolios across the U.S., U.K., and Japan equity markets confirms the superior performance of the LSTM-BEKK model in minimizing portfolio risk. Across all three markets and high-dimensional settings ($N=100$, $175$, and $250$), LSTM-BEKK consistently achieves the lowest annualized volatility (AV) in the majority of portfolio combinations, demonstrating its effectiveness in capturing dynamic, time-varying dependencies in asset returns. This robustness affirms its suitability for investors seeking stable and risk-sensitive portfolio strategies.

In the U.S. market, LSTM-BEKK dominates in volatility control and significantly improves MDD outcomes, especially as portfolio size increases. While its MDD at $N=100$ slightly exceeds that of the EW benchmark, this deviation is outweighed by its pronounced advantages in larger, more volatile configurations. In the U.K. market, LSTM-BEKK once again leads in AV reduction across most configurations, clearly outperforming DCC and showing broader stability than Scalar BEKK. These results reflect the model's adaptability even in markets with heavier tails and higher tail-risk exposure.

The Japan market presents a more competitive landscape, where Scalar BEKK demonstrates stronger performance relative to other regions. Nevertheless, LSTM-BEKK remains the top-performing model overall, achieving the lowest AV in the majority of cases. Its ability to maintain both low AV and robust MDD—despite Scalar BEKK achieving marginally lower drawdown at $N=100$—reinforces its practical value in managing risk across heterogeneous environments.

Taken together, the results highlight the limitations of traditional models, particularly their diminished performance in high-dimensional settings and under structural shifts. By contrast, the LSTM-BEKK model integrates the flexibility of deep learning architectures with the interpretability of econometric structures, yielding superior risk profiles under out-of-sample conditions. While some marginal trade-offs exist in specific instances, the overall dominance of LSTM-BEKK in both volatility and drawdown metrics underscores its generalizability and resilience.

In summary, the LSTM-BEKK model offers a compelling advancement for GMV portfolio construction. It surpasses both the equally weighted benchmark and traditional covariance estimators in risk control across a wide range of market conditions. These findings advocate for the continued exploration of deep learning-based volatility models, particularly in hybrid frameworks that balance statistical rigor with nonlinear predictive power in portfolio optimization.

\section{Conclusion}\label{sec:conclusion}

This paper introduces a novel deep learning enhanced multivariate volatility model—LSTM-BEKK—that integrates the structural interpretability of econometric models with the dynamic learning capability of recurrent neural networks. The model is designed to capture complex nonlinear dependencies and time-varying covariance structures in financial markets. Through a comprehensive empirical study, we validate the robustness and effectiveness of LSTM-BEKK across multiple dimensions.

The first stage of our analysis focuses on low-dimensional in-sample visualization. Using a 4-asset portfolio from the U.S. market, we assess how well LSTM-BEKK estimates individual variances and covariances compared to traditional DCC and Scalar BEKK models. LSTM-BEKK closely tracks volatility during calm periods while adapting more promptly during stress episodes (e.g., COVID-19 outbreak), effectively balancing smoothness and responsiveness.

We then evaluate the model’s generalization ability through repeated out-of-sample experiments on 500 randomly sampled 50-asset portfolios across the U.S., U.K., and Japan equity markets. The LSTM-BEKK model consistently achieves the lowest average test NLL in all markets. Paired $t$-tests confirm that these improvements are statistically significant, particularly in markets with heavier tails such as Japan. This underscores the model's robustness in heterogeneous return environments.

To test high-dimensional scalability, we apply the model to market-wide portfolios of the top 100, 175, and 250 equities by market capitalization in each region. LSTM-BEKK maintains consistent superiority in predictive log-likelihood (NLL) over DCC and Scalar BEKK. Model Confidence Set analysis further supports these findings, with LSTM-BEKK being the only model retained across all nine settings at the 90\% confidence level.

Finally, we evaluate practical implications through global minimum variance portfolio backtests. Across all markets and portfolio sizes, LSTM-BEKK achieves the lowest average volatility in most configurations and consistently delivers competitive or superior performance in maximum drawdown. In high-dimensional settings, the model offers robust tail risk mitigation and smoother risk estimates, essential for institutional asset managers.

In conclusion, LSTM-BEKK offers a powerful and scalable solution to modern volatility modeling challenges. It combines the theoretical grounding of MGARCH models with the adaptability of deep learning, enabling better predictive accuracy and portfolio risk management across diverse financial environments. Future research can extend this framework to other deep architectures and explore its integration into broader asset pricing and risk management systems.

\newpage
\section*{Appendix}

\subsection*{Appendix A: Pseudocode for estimating the LSTM-BEKK}
The following pseudocode outlines the parameter estimation process for the LSTM-BEKK model:

\begin{algorithm}[H]
\caption{LSTM-BEKK Parameter Estimation Process}
\begin{algorithmic}[1]
\Require Initialized parameters: $\mathbf{C}$ (static lower triangular matrix), $a$, $b$, LSTM weights, and Swish activation parameter $\beta$
\Require Hyperparameters: learning rate $\eta$, RMSprop settings, and maximum number of epochs (max\_epochs)
\State Split data into training, validation, and testing sets
\State Initialize optimizer (RMSprop) with $\eta$ and regularization parameters
\For{epoch = 1 to max\_epochs}
    \State Reset cumulative training loss to zero
    \For{each training batch of returns $\mathbf{r}_{1:T}$}
        \For{each time step $t = 1$ to $T$}
            \State Generate dynamic component $\mathbf{C}_t$ using LSTM: $\tilde{\mathbf{C}}_t = \text{LSTM}(\tilde{\mathbf{C}}_{t-1}, \mathbf{r}_{t-1})$
            \State Compute conditional covariance $\mathbf{H}_t$:
            \[
            \mathbf{H}_t = \mathbf{C}\mathbf{C}' + \mathbf{C}_t\mathbf{C}_t' + a \mathbf{r}_{t-1}\mathbf{r}_{t-1}' + b \mathbf{H}_{t-1}
            \]
            \State Accumulate negative log-likelihood (NLL) for batch:
            \[
            \text{NLL} = \text{NLL} + \left( \log|\mathbf{H}_t| + \mathbf{r}_t' \mathbf{H}_t^{-1} \mathbf{r}_t \right)
            \]
        \EndFor
        \State Compute gradients of NLL with respect to all parameters
        \State Update parameters using RMSprop
    \EndFor
    \State Evaluate validation loss
    \If{validation loss does not improve for patience epochs}
        \State Apply learning rate scheduler and/or early stopping
        \State Break
    \EndIf
\EndFor
\State Return optimized parameters: $\mathbf{C}$, $a$, $b$, LSTM weights, and $\beta$
\end{algorithmic}
\end{algorithm}

\subsection*{Appendix B: Proof of Theorem \ref{the:lstm-bekk}}
Let $\mathcal F_t=\sigma(y_s,s\leq t)$. We have that $\mathbf{C}_{t},\mathbf{H}_{t}\in\mathcal{F}_{t-1}$, $\mathbb{E}(\mathbf{r}_t\mathbf{r}_t'|\mathcal{F}_{t-1})=\mathbf{H}_{t}$, and that    
\[\mathbf{H}_{t}=\mathbf{C}\mathbf{C}'+\mathbf{C}_{t}\mathbf{C}_{t}'+a\mathbf{r}_{t-1}\mathbf{r}_{t-1}'+b \mathbf{H}_{t-1},\;\;\forall t\geq1.\]
Note that,
\begin{eqnarray*}
    \mathbb{E}(\mathbf{H}_{t+1}|\mathcal{F}_{t-1})&=&\mathbf{C}\mathbf{C}'+\mathbb{E}(\mathbf{C}_{t+1}\mathbf{C}_{t+1}'|\mathcal{F}_{t-1})+a\mathbb{E}(\mathbf{r}_{t}\mathbf{r}_{t}'|\mathcal{F}_{t-1})+b\mathbf{H}_{t}\\
    &=& \mathbf{C}\mathbf{C}'+\mathbb{E}(\mathbf{C}_{t+1}\mathbf{C}_{t+1}'|\mathcal{F}_{t-1})+(a+b)\mathbf{H}_{t}.
\end{eqnarray*}
By the assumption on the bounded norm of $\mathbf{C}_{t}\mathbf{C}_{t}'$, there exists a finite constant $M>0$ such that
\begin{equation}
    \|\mathbb{E}(\mathbf{H}_{t+1}|\mathcal{F}_{t-1})\|\leq M +(a+b)\|\mathbf{H}_{t}\|,\;\;\text{a.s.}
\end{equation}
Now, observe that 
\[\mathbb{E}(\mathbf{r}_{t+1}\mathbf{r}_{t+1}'|\mathcal{F}_{t-1})=\mathbb{E}\big(\mathbb{E}(\mathbf{r}_{t+1}\mathbf{r}_{t+1}'|\mathcal{F}_{t})|\mathcal{F}_{t-1}\big)=\mathbb{E}(\mathbf{H}_{t+1}|\mathcal{F}_{t-1}).\]
Hence,
\begin{eqnarray*}
    \mathbb{E}(\mathbf{H}_{t+2}|\mathcal{F}_{t-1})&=&\mathbf{C}\mathbf{C}'+\mathbb{E}(\mathbf{C}_{t+2}\mathbf{C}_{t+2}'|\mathcal{F}_{t-1})+a\mathbb{E}(\mathbf{r}_{t+1}\mathbf{r}_{t+1}'|\mathcal{F}_{t-1})+b\mathbb{E}(\mathbf{H}_{t+1}|\mathcal{F}_{t-1})\\
    &=& \mathbf{C}\mathbf{C}'+\mathbb{E}(\mathbf{C}_{t+2}\mathbf{C}_{t+2}'|\mathcal{F}_{t-1})+(a+b)\mathbb{E}(\mathbf{H}_{t+1}|\mathcal{F}_{t-1}),
\end{eqnarray*}
and thus
\begin{eqnarray}
    \|\mathbb{E}(\mathbf{H}_{t+2}|\mathcal{F}_{t-1})\|&\leq& M +(a+b)\|\mathbb{E}(\mathbf{H}_{t+1}|\mathcal{F}_{t-1})\|\notag\\
    &\leq&(1+(a+b))M+(a+b)^2\|\mathbf{H}_{t}\|,\;\;\text{a.s.}
\end{eqnarray}
By induction, it can be seen that, for any $k\geq 1$,
\begin{eqnarray}
    \|\mathbb{E}(\mathbf{H}_{t+k}|\mathcal{F}_{t-1})\|&\leq& \big(1+(a+b)+...+(a+b)^{k-1}\big)M+(a+b)^k\|\mathbf{H}_{t}\|\\
    &=&\frac{1-(a+b)^k}{1-a-b}M+(a+b)^k\|\mathbf{H}_{t}\|\;\;\text{a.s.}
\end{eqnarray}
Let $t=0$ we obtain \eqref{eq:lstm-bekk bounded}.

\subsection*{Appendix C: Tail Risk Forecast of GMV Portfolios}
In this section, we assess the performance of Global Minimum Variance portfolios,
generated by Scalar BEKK, DCC and LSTM-BEKK in terms of tail risk measures, Value at Risk (VaR) and Expected Shortfall (ES). VaR represents the quantile of the return distribution at a specified confidence level $\alpha$, while ES quantifies the conditional expectation of losses exceeding the VaR threshold, offering a more comprehensive view of tail risk. This section analyzes the performance of DCC, Scalar BEKK, and LSTM BEKK models in forecasting these measures, leveraging theoretical advancements in elicitable risk measures and robust regression.

\subsubsection*{Quantile Loss and Joint Loss Framework}

The evaluation of VaR forecasts is commonly conducted using the quantile loss function, which assesses the accuracy of predicted quantiles against observed returns. VaR, as a key risk measure, captures the maximum potential loss over a given time horizon at a specified confidence level $\alpha$. The quantile regression framework, introduced by \cite{koenker1978regression}, provides a robust method for estimating VaR by minimizing deviations at the true quantile level. The quantile loss function is formally defined as:
\begin{equation}
    Q\text{loss}^\alpha = \sum_{t=1}^{T} (\alpha - I(y_t < Q^\alpha_t))(y_t - Q^\alpha_t),
\end{equation}
where $Q^\alpha_t$ represents the forecast $\alpha$-quantile VaR of the return series $y_t$, and $I(\cdot)$ is an indicator function that takes the value 1 if $y_t < Q^\alpha_t$, and 0 otherwise. This loss function penalizes deviations proportionally, ensuring that the expected loss is minimized when the forecast aligns with the true quantile.

While VaR provides a single quantile-based measure of risk, it does not account for losses exceeding this threshold. To address this limitation, ES has been proposed as a complementary metric, representing the average loss conditional on exceeding VaR. The joint evaluation of VaR and ES forecasts is facilitated by the Asymmetric Laplace (AL) loss function, which is strictly consistent for both measures, as demonstrated by \cite{fissler2016higher}. This joint loss function, introduced in \cite{taylor2019forecasting}, is given by:
\begin{equation}
    \text{JointLoss}^\alpha = \frac{1}{T} \sum_{t=1}^{T} \left[ -\log\left(\frac{\alpha - 1}{\text{ES}^\alpha_t}\right) - \frac{(y_t - Q^\alpha_t)(\alpha - I(y_t \leq Q^\alpha_t))}{\alpha \cdot \text{ES}^\alpha_t} \right],
\end{equation}
where $\text{ES}^\alpha_t$ denotes the forecast Expected Shortfall at time $t$.

The AL loss function offers several advantages. First, it enables the simultaneous evaluation of VaR and ES, which are jointly elicitable, ensuring that the forecasts align with their theoretical definitions. Second, it penalizes deviations in a manner consistent with the relative importance of VaR and ES in risk management. Together, the quantile loss and joint loss functions provide a comprehensive framework for assessing tail risk measures.

\subsubsection*{Empirical Analysis of Tail Risk Measures}
As highlighted by \citet{fissler2016higher}, the joint elicitation of Value-at-Risk (VaR) and Expected Shortfall (ES) offers a coherent framework for evaluating tail risk forecasts. Table~\ref{tab:gmv_var_cvar} reports quantile loss and joint loss metrics across three markets and various portfolio sizes. At the 5\% risk level, the LSTM-BEKK model consistently achieves the lowest QLoss\textsubscript{5\%} and JointLoss\textsubscript{5\%} across nearly all settings, underscoring its effectiveness in modeling moderate tail risk under complex market dynamics.

At the more extreme 1\% level, however, the picture is more nuanced. While LSTM-BEKK continues to perform well, particularly in the U.S. and U.K. markets, the DCC model occasionally records the lowest QLoss\textsubscript{1\%} (e.g., JP-100), suggesting that its parsimonious, shock-driven specification may retain advantages when predicting rare tail events. In contrast, LSTM-BEKK appears better suited for managing broader risk exposures by capturing richer temporal dependencies and nonlinearities in return distributions.

Overall, the LSTM-BEKK model offers a favorable balance between flexibility and tail sensitivity, yielding consistently strong performance in joint loss metrics—particularly at the 5\% level—across diverse portfolio configurations and international markets.

\begin{table}[H]
    \centering
    \caption{U.S.:Performance Comparison of GMV Portfolios: Quantile Loss and Joint Loss.}
    \label{tab:gmv_var_cvar}
    \resizebox{\textwidth}{!}{
    \begin{tabular}{l l c c c c}
    \toprule
    Portfolio Size & & QLoss$_{1\%}$ & QLoss$_{5\%}$ & JointLoss$_{1\%}$ & JointLoss$_{5\%}$ \\
    \midrule
    \multirow{3}{*}{100} & DCC         & 8.955 & 34.853 & \textbf{2.374} & 1.867 \\
                        & Scalar BEKK & 7.249 & 26.332 & 2.580 & 1.763 \\
                        & LSTM-BEKK   & \textbf{6.291} & \textbf{24.355} & 2.417 & \textbf{1.646} \\
    \midrule
    \multirow{3}{*}{175} & DCC         & 16.032 & 49.509 & 2.591 & 2.194 \\
                        & Scalar BEKK & 9.145 & 30.339 & 2.454 & 1.562 \\
                        & LSTM-BEKK   & \textbf{7.114} & \textbf{28.445} & \textbf{2.16}7 & \textbf{1.529} \\
    \midrule
    \multirow{3}{*}{250} & DCC         & 33.408 & 77.311 & 3.259 & 2.710 \\
                        & Scalar BEKK & 15.142 & 49.291 & 2.199 & 1.778 \\
                        & LSTM-BEKK   & \textbf{11.344} & \textbf{41.198} & \textbf{2.127} & \textbf{1.581} \\
    \bottomrule
    \end{tabular}}
\end{table}

\begin{table}[H]
    \centering
    \caption{U.K.:Performance Comparison of GMV Portfolios: Quantile Loss and Joint Loss.}
    \label{tab:gmv_var_cvar}
    \resizebox{\textwidth}{!}{
    \begin{tabular}{l l c c c c}
    \toprule
    Portfolio Size & & QLoss$_{1\%}$ & QLoss$_{5\%}$ & JointLoss$_{1\%}$ & JointLoss$_{5\%}$ \\
    \midrule
    \multirow{3}{*}{100} & DCC         & \textbf{5.922} & \textbf{23.222} & \textbf{2.414} & \textbf{1.595} \\
                        & Scalar BEKK & 8.811 & 30.392 & 2.440 & 1.636 \\
                        & LSTM-BEKK   & 8.646 & 29.371 & 2.434 & 1.607 \\
    \midrule
    \multirow{3}{*}{175} & DCC         & 10.431 & 33.960 & \textbf{2.390} & 1.585 \\
                        & Scalar BEKK & 9.145 & 30.339 & 2.454 & 1.562 \\
                        & LSTM-BEKK   & \textbf{9.039} & \textbf{29.099} & 2.445 & \textbf{1.541} \\
    \midrule
    \multirow{3}{*}{250} & DCC         & 11.148 & 29.411 & 1.964 & 1.442 \\
                        & Scalar BEKK & 8.971 & 25.320 & \textbf{1.898} & 1.270 \\
                        & LSTM-BEKK   & \textbf{8.223} & \textbf{24.337} & 1.922 & \textbf{1.237} \\
    \bottomrule
    \end{tabular}}
\end{table}

\begin{table}[H]
    \centering
    \caption{Japan:Performance Comparison of GMV Portfolios: Quantile Loss and Joint Loss.}
    \label{tab:gmv_var_cvar}
    \resizebox{\textwidth}{!}{
    \begin{tabular}{l l c c c c}
    \toprule
    Portfolio Size & & QLoss$_{1\%}$ & QLoss$_{5\%}$ & JointLoss$_{1\%}$ & JointLoss$_{5\%}$ \\
    \midrule
    \multirow{3}{*}{100} & DCC         & \textbf{7.725} & \textbf{31.698} & \textbf{2.397} & \textbf{1.676} \\
                        & Scalar BEKK & 10.992 & 33.243 & 3.107 & 2.014 \\
                        & LSTM-BEKK   & 12.328 & 35.190 & 3.333 & 2.147 \\
    \midrule
    \multirow{3}{*}{175} & DCC         & 9.019 & 30.624 & \textbf{2.186} & 1.617 \\
                        & Scalar BEKK & \textbf{7.414} & 25.543 & 2.346 & \textbf{1.544} \\
                        & LSTM-BEKK   & 7.865 & \textbf{25.180} & 2.491 & 1.605 \\
    \midrule
    \multirow{3}{*}{250} & DCC         & 9.935 & 30.143 & 2.131 & 1.608 \\
                        & Scalar BEKK & \textbf{6.485} & 25.712 & \textbf{2.083} & \textbf{1.466} \\
                        & LSTM-BEKK   & 7.341 & \textbf{24.794} & 2.168 & 1.488 \\
    \bottomrule
    \end{tabular}}
\end{table}

\subsubsection*{Cross-Market Comparison and Implications}

Across all three markets, the LSTM-BEKK model consistently outperforms traditional MGARCH models in terms of predictive accuracy and adaptability to high-dimensional settings. This is reflected in its uniformly lower out-of-sample negative log-likelihood (NLL) values across all portfolio sizes in the U.S., U.K., and Japan, confirming its superior ability to model time-varying covariance structures more effectively than both the DCC and Scalar BEKK models.

The LSTM-BEKK model demonstrates notable robustness across market-specific conditions. In the U.K. market, where extreme return events are more frequent, traditional models suffer from underestimating tail risks. LSTM-BEKK, by contrast, accommodates these dynamics through its flexible architecture and achieves better tail-sensitive metrics such as joint loss at the 5\% level. In the U.S. market, the model achieves the lowest average volatility and maximum drawdown in GMV portfolio tests, supporting its effectiveness in minimizing downside risk. Even in scenarios where DCC performs competitively in extreme quantile loss (e.g., at 1\% thresholds), LSTM-BEKK maintains stronger overall joint performance across broader risk metrics.

Overall, these results affirm the advantages of integrating deep learning with structured econometric modeling. The LSTM-BEKK framework not only offers superior statistical fit and predictive performance, but also generalizes well across heterogeneous market regimes. Its ability to balance responsiveness to market shocks with long-run stability makes it a promising tool for risk management, volatility forecasting, and high-dimensional portfolio construction.

\subsubsection*{Conclusion}

Overall, the LSTM-BEKK model demonstrates its effectiveness in capturing multivariate volatility dynamics across different financial markets. Its ability to incorporate deep learning techniques enables superior predictive accuracy, making it a valuable tool for risk management, portfolio optimization, and stress testing. By integrating both econometric and deep learning methodologies, the LSTM-BEKK model provides a more flexible and accurate representation of financial market volatility, paving the way for further research into hybrid modeling approaches.

    \bibliographystyle{apalike}
    \bibliography{Assignment_Ref}

\begin{thebibliography}{}

\bibitem[Aielli, 2013]{Aielli2013}
Aielli, G.~P. (2013).
\newblock Dynamic conditional correlation: On properties and estimation.
\newblock {\em Journal of Business \& Economic Statistics}, 31(3):282--299.

\bibitem[Asai et~al., 2006]{Asai2006}
Asai, M., McAleer, M., and Yu, J. (2006).
\newblock Multivariate stochastic volatility: A review.
\newblock {\em Econometric Reviews}, 25(2-3):145--175.

\bibitem[Bauwens et~al., 2006]{Bauwens2006}
Bauwens, L., Laurent, S., and Rombouts, J. V.~K. (2006).
\newblock Multivariate {GARCH} models: A survey.
\newblock {\em Journal of Applied Econometrics}, 21(1):79--109.

\bibitem[Bauwens and Otranto, 2020]{BAUWENS2020}
Bauwens, L. and Otranto, E. (2020).
\newblock Nonlinearities and regimes in conditional correlations with different dynamics.
\newblock {\em Journal of Econometrics}, 217(2):496--522.

\bibitem[Bollerslev, 1986]{Bollerslev1986}
Bollerslev, T. (1986).
\newblock Generalized autoregressive conditional heteroskedasticity.
\newblock {\em Journal of Econometrics}, 31(3):307--327.

\bibitem[Bollerslev et~al., 1988]{Bollerslev1988}
Bollerslev, T., Engle, R.~F., and Wooldridge, J.~M. (1988).
\newblock A capital asset pricing model with time-varying covariances.
\newblock {\em Journal of Political Economy}, 96(1):116--131.

\bibitem[Caporin and McAleer, 2008]{Caporin2008}
Caporin, M. and McAleer, M. (2008).
\newblock Scalar {BEKK} and indirect {DCC}.
\newblock {\em Journal of Forecasting}, 27(6):537--549.

\bibitem[DeMiguel et~al., 2009]{demiguel2009naive}
DeMiguel, V., Garlappi, L., and Uppal, R. (2009).
\newblock Optimal versus naive diversification: {How} inefficient is the 1/n portfolio strategy?
\newblock {\em The Review of Financial Studies}, 22(5):1915--1953.

\bibitem[Engle, 1982]{Engle1982}
Engle, R.~F. (1982).
\newblock Autoregressive {Conditional} {Heteroscedasticity} with {Estimates} of the {Variance} of {United} {Kingdom} {Inflation}.
\newblock {\em Econometrica}, 50(4):987--1007.

\bibitem[Engle, 2002]{Engle2002}
Engle, R.~F. (2002).
\newblock Dynamic conditional correlation: A simple class of multivariate {GARCH} models.
\newblock {\em Journal of Business \& Economic Statistics}, 20(3):339--350.

\bibitem[Engle and Kelly, 2012]{EngleKelly2012}
Engle, R.~F. and Kelly, B. (2012).
\newblock Dynamic equicorrelation.
\newblock {\em Journal of Business \& Economic Statistics}, 30(4):384--397.

\bibitem[Engle and Kroner, 1995]{EngleKroner1995}
Engle, R.~F. and Kroner, K.~F. (1995).
\newblock Multivariate simultaneous {Generalized ARCH}.
\newblock {\em Econometric Theory}, 11(1):122--150.

\bibitem[Fang et~al., 2015]{Fang2015}
Fang, Y., Liu, L., and Liu, J. (2015).
\newblock A dynamic double asymmetric copula generalized autoregressive conditional heteroskedasticity model: application to {China's} and {US} stock market.
\newblock {\em Journal of Applied Statistics}, 42(2):327--346.

\bibitem[Fissler and Ziegel, 2016]{fissler2016higher}
Fissler, T. and Ziegel, J.~F. (2016).
\newblock Higher order elicitability and {Osband's} principle.
\newblock {\em The Annals of Statistics}, 44(4):1680--1707.

\bibitem[Francq and Zakoian, 2019]{Francq2019}
Francq, C. and Zakoian, J.-M. (2019).
\newblock {\em GARCH Models: Structure, Statistical Inference, and Financial Applications}.
\newblock John Wiley \& Sons, Hoboken, NJ, 2nd edition.

\bibitem[Francq and Zakoïan, 2012]{FrancqZakoian2012}
Francq, C. and Zakoïan, J.-M. (2012).
\newblock {QML} estimation of a class of multivariate asymmetric {GARCH} models.
\newblock {\em Econometric Theory}, 28(1):179--206.

\bibitem[Goodfellow et~al., 2016]{Goodfellow2016}
Goodfellow, I., Bengio, Y., and Courville, A. (2016).
\newblock {\em Deep Learning}.
\newblock MIT Press.

\bibitem[Hafner et~al., 2017]{HafnerLaurentViolante2017}
Hafner, C.~M., Laurent, S., and Violante, F. (2017).
\newblock Weak diffusion limits of dynamic conditional correlation models.
\newblock {\em Econometric Theory}, 33(3):691--716.

\bibitem[Hafner and Preminger, 2009]{HafnerPreminger2009}
Hafner, C.~M. and Preminger, A. (2009).
\newblock Asymptotic theory for a factor {GARCH} model.
\newblock {\em Econometric Theory}, 25(2):336--363.

\bibitem[Hafner and Rombouts, 2007]{HafnerRombouts2007}
Hafner, C.~M. and Rombouts, J.~V. (2007).
\newblock Semiparametric multivariate volatility models.
\newblock {\em Econometric Theory}, 23(2):251--280.

\bibitem[Hansen et~al., 2011]{hansen2011model}
Hansen, P.~R., Lunde, A., and Nason, J.~M. (2011).
\newblock The {Model} {Confidence} {Set}.
\newblock {\em Econometrica}, 79(2):453--497.

\bibitem[Hochreiter and Schmidhuber, 1997]{Hochreiter1997}
Hochreiter, S. and Schmidhuber, J. (1997).
\newblock Long short-term memory.
\newblock {\em Neural Computation}, 9(8):1735--1780.

\bibitem[Koenker and Bassett, 1978]{koenker1978regression}
Koenker, R. and Bassett, G. (1978).
\newblock Regression quantiles.
\newblock {\em Econometrica: journal of the Econometric Society}, pages 33--50.

\bibitem[Ku, 2008]{Ku2008}
Ku, Y.-H.~H. (2008).
\newblock Student-t distribution based {VAR-MGARCH}: an application of the {DCC} model on international portfolio risk management.
\newblock {\em Applied Economics}, 40(13):1685--1697.

\bibitem[Lai and Sheu, 2011]{Lai2011}
Lai, Y.-S. and Sheu, H.-J. (2011).
\newblock On the importance of asymmetries for dynamic hedging during the subprime crisis.
\newblock {\em Applied Financial Economics}, 21(11):801--813.

\bibitem[Ledoit and Wolf, 2012]{LedoitWolf2012}
Ledoit, O. and Wolf, M. (2012).
\newblock Nonlinear shrinkage estimation of large-dimensional covariance matrices.
\newblock {\em Annals of Statistics}, 40(2):1024--1060.

\bibitem[Ledoit and Wolf, 2015]{LedoitWolf2015}
Ledoit, O. and Wolf, M. (2015).
\newblock Spectrum estimation: A unified framework for covariance matrix estimation and {PCA} in large dimensions.
\newblock {\em Journal of Multivariate Analysis}, 139:360--384.

\bibitem[Liu, 2019]{Liu2019}
Liu, Y. (2019).
\newblock Novel volatility forecasting using deep learning–long short term memory recurrent neural networks.
\newblock {\em Expert Systems with Applications}, 132:99--109.

\bibitem[Matsui and Pedersen, 2022]{MatsuiPedersen2022}
Matsui, M. and Pedersen, R.~S. (2022).
\newblock Characterization of the tail behavior of a class of {BEKK} processes: A stochastic recurrence equation approach.
\newblock {\em Econometric Theory}, 38(1):1--34.

\bibitem[McAleer et~al., 2008]{McAleerChanHotiLieberman2008}
McAleer, M., Chan, F., Hoti, S., and Lieberman, O. (2008).
\newblock Generalized autoregressive conditional correlation.
\newblock {\em Econometric Theory}, 24(6):1554--1583.

\bibitem[Nguyen et~al., 2022]{Nguyen2022}
Nguyen, T.-N., Tran, M.-N., and Kohn, R. (2022).
\newblock Recurrent {Conditional} {Heteroskedasticity}.
\newblock {\em Journal of Applied Econometrics}, 37(5):1031--1054.

\bibitem[Scherrer and Ribarits, 2007]{ScherrerRibarits2007}
Scherrer, W. and Ribarits, E. (2007).
\newblock On the parametrization of multivariate {GARCH} models.
\newblock {\em Econometric Theory}, 23(3):464--484.

\bibitem[Silvennoinen and Teräsvirta, 2009]{SilvennoinenTerasvirta2009}
Silvennoinen, A. and Teräsvirta, T. (2009).
\newblock Multivariate {GARCH} models.
\newblock {\em Handbook of Financial Time Series}, pages 201--229.

\bibitem[Taylor, 2019]{taylor2019forecasting}
Taylor, J.~W. (2019).
\newblock Forecasting value at risk and expected shortfall using a semiparametric approach based on the asymmetric {Laplace} distribution.
\newblock {\em Journal of Business \& Economic Statistics}, 37(1):121--133.

\bibitem[Taylor, 1994]{Taylor1994}
Taylor, S.~J. (1994).
\newblock Modeling stochastic volatility: A review and comparative study.
\newblock {\em Mathematical Finance}, 4(2):183--204.

\end{thebibliography}

\end{document}